\documentclass[twocolumn,floatfix,aps,tightenlines,prd,nofootinbib,showpacs,superscriptaddress]{revtex4}
\usepackage{amsmath,amssymb,amsfonts}
\usepackage{graphicx}
\usepackage{psfrag}
\usepackage{dcolumn}
\usepackage{bm}
\usepackage{comment}

\usepackage[usenames]{color}
\usepackage{latexsym}

\newcommand {\be}{\begin{equation}}
\newcommand {\ee}{\end{equation}}
\newcommand {\ba}{\begin{array}}
\newcommand {\ea}{\end{array}}
\newcommand {\bea}{\begin{eqnarray}}
\newcommand {\eea}{\end{eqnarray}}
\newcommand {\bean}{\begin{eqnarray*}}
\newcommand {\eean}{\end{eqnarray*}}
\newcommand{\etal}{{\it et al.~}\!}

\newcommand {\vecnorm}[2]{\parallel \vec{#1}_{#2} \parallel}
\newcommand {\vecdotvec}[2]{\vec{#1} \cdot \vec{#2}}
\newcommand {\vecdothat}[2]{\vec{#1} \cdot \hat{#2}}

\newcommand {\mI}{\mathcal{I}}
\newcommand {\mJ}{\mathcal{J}}
\newcommand {\mS}{\mathcal{S}}
\newcommand {\mM}{\mathcal{M}}

\newcommand {\cth}{\cos\theta}
\newcommand {\sth}{\sin\theta}
\newcommand {\cph}{\cos\phi}
\newcommand {\sph}{\sin\phi}
\newcommand {\ccth}{\cos^2\theta}

\newcommand {\defn}{\mathrel{\mathop:}=}

\begin{document}

\title{Black hole puncture initial data with realistic gravitational wave content}

\author{B.~J.~Kelly}
\affiliation{Gravitational Astrophysics Laboratory, 
NASA Goddard Space Flight Center, 
8800 Greenbelt Rd., Greenbelt, MD 20771, USA}
\affiliation{Center for Gravitational Wave Astronomy, 
Department of Physics and Astronomy,
The University of Texas at Brownsville, Brownsville, Texas 78520}
\author{W.~Tichy} \affiliation{Department of Physics, Florida Atlantic University, Boca Raton Florida 33431-0991}
\author{M.~Campanelli} 
\affiliation{Center for Computational Relativity and Gravitation,
School of Mathematical Sciences, Rochester Institute of Technology, 
78 Lomb Memorial Drive, Rochester,
 New York 14623}
\affiliation{Center for Gravitational Wave Astronomy, 
Department of Physics and Astronomy,
The University of Texas at Brownsville, Brownsville, Texas 78520}
\author{B.~F.~Whiting} \affiliation{Department of Physics, University of Florida, Gainsville Florida 32611-8440}
\affiliation{Center for Gravitational Wave Astronomy, 
Department of Physics and Astronomy,
The University of Texas at Brownsville, Brownsville, Texas 78520}

\date{\today}

\begin{abstract}
  We present improved post-Newtonian-inspired initial data for
  non-spinning black-hole binaries, suitable for numerical evolution
  with punctures. We revisit the work of Tichy \etal [W. Tichy,
  B. Br\"ugmann, M. Campanelli, and P. Diener, Phys. Rev. D {\bf 67},
  064008 (2003)], explicitly calculating the remaining integral terms.
  These terms improve accuracy in the far zone and, for the first
  time, include realistic gravitational waves in the initial data.  We
  investigate the behavior of these data both at the center of mass
  and in the far zone, demonstrating agreement of the
  transverse-traceless parts of the new metric with
  quadrupole-approximation waveforms. These data can be used for
  numerical evolutions, enabling a direct connection between the
  merger waveforms and the post-Newtonian inspiral waveforms.

\end{abstract}
\maketitle

\section{Introduction}\label{Sec:intro}

Post-Newtonian (PN) methods have played a fundamental role in our
understanding of the astrophysical implications of Einstein's theory
of general relativity.  Most importantly, they have been used to
confirm that the radiation of gravitational waves accounts for energy
loss in known binary pulsar configurations.  They have also been used
to create templates for the gravitational waves emitted from compact
binaries which might be detected by ground-based gravitational wave
observatories, such as LIGO~\cite{LIGO,Abbott:2003vs}, and the
NASA/ESA planned space-based mission,
LISA~\cite{LISA:1998pa,Danzmann:2003tv}.  However, PN methods have not
been extensively used to provide initial data for binary evolution in
numerical relativity, nor, until recently (see
\cite{Buonanno:2006ui,Berti:2007fi}), have they been extensively studied so that
their limitations could be well identified and the results of
numerical relativity independently confirmed.

Until the end of 2004, the field of numerical relativity had been
struggling to compute even a single orbit for a black-hole binary
(BHB). Although debate occurred on the advantages of one type of
initial data over another, the primary focus within the numerical
relativity community was on code refinement which would lead to more
stable evolution. Astrophysical realism was very much a secondary
issue.  However, this situation has radically changed in the last few
years with the introduction of two essentially independent, but
equally successful techniques: the generalized harmonic gauge (GHG)
method developed by Pretorius ~\cite{Pretorius:2005gq} and the
``moving puncture'' approach, independently developed by the UTB and
NASA Goddard groups \cite{Campanelli:2005dd,Baker:2005vv}.  Originally
introduced by Brandt \& Br\"{u}gmann \cite{Brandt:1997tf} in the
context of initial data, the puncture method explicitly factored out
the singular part of the metric.  When used in numerical evolution in
which the punctures remained fixed on the numerical grid, it resulted
in distortions of the coordinate system and instabilities in the
Baumgarte-Shapiro-Shibata-Nakamura (BSSN) \cite{Shibata:1995we,Baumgarte:1998te}
evolution scheme.  The revolutionary idea behind the moving puncture
approach was precisely, not to factor out the singular part of the
metric, but rather evolve it together with the regular part, allowing
the punctures to move freely across the grid with a suitable choice of
the gauge.

A golden age for numerical relativity is now emerging, in which
multiple groups are using different computer codes to evolve BHBs for
several orbits before plunge and merger
\cite{Bruegmann:2003aw,Campanelli:2006gf,Pretorius:2006tp,Baker:2006yw,
  Bruegmann:2006at,Scheel:2006gg,Marronetti:2007ya,Tichy:2006qn,Pfeiffer:2007yz}.
Comparison of the numerical results obtained from these various codes
has taken place ~\cite{Baker:2007fb,Thornburg:2007hu,NRwaves}, and
comparison with PN inspiral waveforms has also been carried out with
encouraging success \cite{Buonanno:2006ui,Berti:2007fi,Baker:2006kr,
  Baker:2006ha}.  The application of successful numerical relativity
tools to study some important astrophysical properties (e.g.\
precession, recoil, spin-orbit coupling, elliptical orbits, etc) of
spinning and/or unequal mass-black hole systems is currently producing
extremely interesting new results ~\cite{Campanelli:2004zw,
  Herrmann:2006ks, Baker:2006vn,Campanelli:2006uy,Campanelli:2006fg,
  Campanelli:2006fy,
  Gonzalez:2006md,Herrmann:2007ac,Campanelli:2007ew,
  Koppitz:2007ev,Gonzalez:2007hi,Choi:2007eu,Baker:2007gi,Pretorius:2007jn,
  Campanelli:2007cg,Tichy:2007hk}.  It now seems that the primary
obstacle to further progress is simply one of computing power.  In
this new situation, it is perhaps time to return to the question of
what initial data will best describe an astrophysical BHB.

To date, the best-motivated description of pre-merger BHBs has been
supplied by PN methods.  We might expect, then, that a PN-based
approach would give us the most astrophysically correct initial data
from which to run full numerical simulations.  In practice, PN results
are frequently obtained in a form ill-adapted to numerical
evolution. PN analysis often deals with the full four-metric, in
harmonic coordinates; numerical evolutions frequently use ADM-type
coordinates, with a canonical decomposition of the four-metric into a
spatial metric and extrinsic curvature.

Fortunately, many PN results have been translated into the language of
ADM by Ohta, Damour, Sch\"{a}fer and collaborators. Explicit results
for 2.5PN BHB data in the near zone were given by Sch\"{a}fer
\cite{Schafer:1986rd} and Jaranowski \& Sch\"{a}fer (JS)
\cite{Jaranowski:1997ky}, and these were implemented numerically by
Tichy \etal \cite{Tichy:2002ec}. Their insight was that the
ADM-transverse-traceless (TT) gauge used by Sch\"{a}fer was
well-adapted to a puncture approach.  To facilitate comparison with
this earlier work \cite{Tichy:2002ec}, we continue to use the results
of Sch\"{a}fer and co-workers, anticipating that higher-order PN
results should eventually become available in a useful form.

The initial data provided previously by Tichy \etal already include PN
information. They are accurate up to order $(v/c)^5$ in the near zone
($r \ll \lambda$), but the accuracy drops to order $(v/c)^3$ in the
far zone ($r \gg \lambda$) [here $\lambda \sim \pi \sqrt{r_{12}^3/G
  (m_1+m_2)}$ is the gravitational wavelength].  These data were
incomplete in the sense that they did not include the correct TT
radiative piece in the metric, and thus did not contain realistic
gravitational waves.

In this paper, we revisit the PN data problem in ADM-TT coordinates,
with the aim of supplying Numerical Relativity with initial BHB data
that extend as far as necessary, and contain realistic gravitational
waves. To do this, we have evaluated the ``missing pieces'' of
Sch\"{a}fer's TT metric for the case of two non-spinning particles. We
have analyzed the near- and far-zone behavior of these data, and
incorporated them numerically in the Cactus \cite{Cactus_webpage} framework.  In principle,
the most accurate PN metric available could be used at this step, but
it is not currently available in ADM-TT form.

The remainder of this paper is laid out as follows. In Section
\ref{Sec:ADM-TT}, we summarize the results of Sch\"{a}fer (1985)
\cite{Schafer:1986rd}, and Jaranowski \& Sch\"{a}fer (1997)
\cite{Jaranowski:1997ky} and their application by Tichy \etal (2003)
\cite{Tichy:2002ec}, to the production of puncture data for numerical
evolution.  In Section \ref{Sec:hTT_calc}, we describe briefly the
additional terms necessary to complete $h^{{\rm TT}}$ to order
$(v/c)^4$, deferring details to the Appendix.  In
Section \ref{Sec:numerics}, we study the full data both analytically
and numerically. Section \ref{Sec:discuss} summarizes our results, and
lays the groundwork for numerical evolution of these data, to be
presented in a subsequent article.

\section{ADM-TT Gauge in Post-Newtonian Data}\label{Sec:ADM-TT}

The ``ADM-TT'' gauge \cite{Ohta:1974kp,Schafer:1986rd} is a 3+1 split
of data where the three-metric differs from conformal flatness
precisely by a TT radiative part:
\bea
g_{ij} &=& \left( 1 + \frac{1}{8} \, \phi \right)^4 \, \eta_{ij} + h^{{\rm TT}}_{ij},\\
\pi^{i}_{i} &=&0.
\eea
The fields $\phi$, $\pi^{ij}$ and $h^{{\rm TT}}_{ij}$ can all be
expanded in a post-Newtonian series. Solving the constraint equations
of 3+1 general relativity in this gauge,
\cite{Schafer:1986rd,Jaranowski:1997ky} obtained explicit expressions
valid up to $O(v/c)^5$ in the near zone, incorporating an arbitrary
number of spinless point particles, with arbitrary masses $m_A$. For
$N$ particles, the lowest-order contribution to the conformal factor
is\footnote{We explicitly include the gravitational constant $G$ in
  all expressions here, as the standard convention $G=1$ used in
  Numerical Relativity differs from the convention $16\pi G = 1$
  employed by \cite{Schafer:1986rd,Jaranowski:1997ky}.}:
\be
\label{eqn:phi_2_sol}
\phi^{(2)} = 4 G \sum_{A=1}^N \frac{m_A}{r_A},
\ee
where $r_A = \sqrt{\vec{x} - \vec{x}_A}$ is the distance from the
field point to the location of particle $A$.

In principle $h^{{\rm TT}}_{i j}$ is computed from
\be
\label{eqn:htt_def}
h^{{\rm TT}}_{ij} = -\delta^{{\rm TT}\, kl}_{ij} \Box_{ret}^{-1} s_{kl} ,
\ee
where $\Box_{ret}^{-1}$ is the (flat space) inverse d'Alembertian
(with a ``no-incoming-radiation'' condition \cite{Fock}), $s_{kl}$ is
a non-local source term and $\delta^{{\rm TT}\, kl}_{ij}$ is the
TT-projection operator.  In order to compute $h^{{\rm TT}}_{ij}$ we
first rewrite Eq.~(\ref{eqn:htt_def}) as
\bea
\label{eqn:htt1}
h^{{\rm TT}}_{ij} 
&=& -\delta^{{\rm TT}\, kl}_{ij} 
   \left[\Delta^{-1} + (\Box_{ret}^{-1} -\Delta^{-1})\right] s_{kl} \nonumber \\
&=& h^{{\rm TT}\,({\rm NZ})}_{ij} -\delta^{{\rm TT}\, kl}_{ij}(\Box_{ret}^{-1} -\Delta^{-1}) s_{kl} .
\eea
\smallskip Note that the near-zone approximation $h^{{\rm TT}\,({\rm
    NZ})}_{ij}$ of $h^{{\rm TT}}_{ij}$ has already been computed
in~\cite{Schafer:1986rd} up to order $O(v/c)^4$ (see also
Eq.~\ref{eqn:JS_htt4} below).  The last term in Eq.~(\ref{eqn:htt1})
is difficult to compute because
\be
s_{kl} =
16\pi G \sum_A \, \frac{p_{Ak} \, p_{Al}}{m_A} \delta(x-x_A)
+ \frac{1}{4} \phi^{(2)}_{,k} \phi^{(2)}_{,l} 
\ee
is a non-local source. However, we can approximate $s_{kl}$ by 
\bea
\bar{s}_{kl} &=& 
\sum_A  \left[ \frac{p_{Ak} \, p_{Al}}{m_A} 
- \frac{G}{2} \, \sum_{B \neq A} \, m_A \, m_B \, \frac{n_{ABk} \, n_{ABl}}{r_{AB}} 
\right] \nonumber \\
&& \times 16\pi G \, \delta(x-x_A) .
\eea
and show that
\be
h^{{\rm TT}}_{ij, (div)} 
=    -\delta^{{\rm TT}\, kl}_{ij} (\Box_{ret}^{-1} -\Delta^{-1}) (s_{kl} - \bar{s}_{kl})
\sim O(v/c)^{5} 
\ee
in the near zone. Furthermore, outside the near zone $h^{{\rm
    TT}}_{ij, (div)} \sim 1/r^2$, so that $h^{{\rm TT}}_{ij, (div)}$
falls off much faster than rest of $h^{{\rm TT}}_{ij}$, which falls
off like $1/r$.  Hence
\be
h^{{\rm TT}}_{ij}
= h^{{\rm TT}\,({\rm NZ})}_{ij} -\delta^{{\rm TT}\, kl}_{ij}(\Box_{ret}^{-1} -\Delta^{-1}) \bar{s}_{kl}
 + h^{{\rm TT}}_{ij, (div)} ,
\ee
where $h^{{\rm TT}}_{ij, (div)}$ can be neglected if we only keep
terms up to $O(v/c)^4$ generally, and $O(1/r)$ at infinity.

The full expression for $h^{{\rm TT}}_{i j}$ for $N$ interacting point
particles from Eq.~(4.3) of \cite{Schafer:1986rd} is:
\bea
\label{eqn:full_htt4}
h^{{\rm TT}}_{i j} &=& h^{{\rm TT}\,({\rm NZ})}_{i j}  + h^{{\rm TT}}_{i j, (div)} 
+ 16\pi\,G \, \int \frac{d^3\vec{k} \, d\omega \, d\tau}{(2 \, \pi)^4}  \nonumber \\
&& \times \sum_A \, \left[ \frac{p_{Ai} \, p_{Aj}}{m_A} - \frac{G}{2} \sum_{B \neq A} \, m_A \, m_B \, \frac{n_{ABi} \, n_{ABj}}{r_{AB}} \right]^{{\rm TT}}_{\tau} \nonumber \\
&& \times \frac{(\omega/k)^2\,e^{i \, \vec{k} \cdot (\vec{x} - \vec{x}_A) - i \, \omega \, (t - \tau)} }{k^2 - (\omega + i \, \epsilon)^2} .
\eea
The first term in (\ref{eqn:full_htt4}), $h^{{\rm TT}\,({\rm NZ})}_{i
  j}$ can be expanded in $v/c$ as
\be
h^{{\rm TT}\,({\rm NZ})}_{ij} = h^{{\rm TT}\,(4)}_{ij} + h^{{\rm TT}\,(5)}_{ij} + O(v/c)^6 .
\ee
The leading order term at $O(v/c)^4$, is given explicitly by Eq.~(A20)
of \cite{Jaranowski:1997ky}:
\begin{widetext}
\bea
\label{eqn:JS_htt4}
h^{{\rm TT}\,(4)ij} &=& \frac{G}{4} \, \sum_A \frac{1}{m_A \, r_A} \left\{ \left[ \vecnorm{p}{A}^2 - 5 \, (\hat{n}_A \cdot \vec{p}_A)^2 \right] \delta^{ij} + 2 \, p_A^i \, p_A^j + \left[ 3 (\hat{n}_A \cdot \vec{p}_A)^2 - 5 \vecnorm{p}{A}^2 \right] n_A^i \, n_A^j + 12 (\hat{n}_A \cdot \vec{p}_A) n_A^{(i} p_A^{j)} \right\} \nonumber \\
&& +  \frac{G^2}{8} \, \sum_A \, \sum_{B \neq A} \, m_A \, m_B \, \left\{ - \frac{32}{s_{AB}} \, \left( \frac{1}{r_{AB}} + \frac{1}{s_{AB}} \right) n^i_{AB} n^j_{AB} + 2 \left(\frac{r_A + r_B}{r^3_{AB}} + \frac{12}{s^2_{AB}} \right) n^i_A \, n^j_B \right. \nonumber \\
&& \left. + 32 \left( \frac{2}{s^2_{AB}} - \frac{1}{r^2_{AB}} \right) n^{(i}_A n^{j)}_{AB} + \left[ \frac{5}{r_{AB} r_A} - \frac{1}{r^3_{AB}} \left( \frac{r^2_B}{r_A} + 3 r_A \right) - \frac{8}{s_{AB}} \left( \frac{1}{r_A} + \frac{1}{s_{AB}} \right) \right] n^i_A n^j_A \right. \nonumber \\
&& \left. + \left[ 5 \frac{r_A}{r^3_{AB}} \left( \frac{r_A}{r_B} - 1 \right) - \frac{17}{r_{AB} r_A } + \frac{4}{r_A r_B} + \frac{8}{s_{AB}} \left( \frac{1}{r_A} + \frac{4}{r_{AB}} \right) \right] \delta^{ij} \right\},
\eea
\end{widetext}
where $s_{AB} \equiv r_A + r_B + r_{AB}$.  The other two terms in
Eq.~(\ref{eqn:full_htt4}) can be shown to be small in the near zone
($r \ll \lambda$, where the characteristic wavelength $\lambda \sim
100M$ for $r_{AB} \sim 10M$).  However, $h^{{\rm TT}\,({\rm NZ})}_{i j}$ is
only a valid approximation to $h^{{\rm TT}}_{i j}$ in the near zone,
and becomes highly inaccurate when used further afield.

Setting aside these far-field issues, Tichy \etal \cite{Tichy:2002ec}
applied Sch\"{a}fer's formulation, in the context of a black-hole
binary system, to construct initial data that are accurate up to
$O(v/c)^5$ in the near zone.  They noted that the ADM-TT decomposition
was well-adapted to the use of a puncture approach to handle
black-hole singularities.  This approach is essentially an extension
of the method introduced in~\cite{Brandt:1997tf}. It allows a simple
numerical treatment of the black holes without the need for excision.

The PN-based puncture data of Tichy {\etal} have not been used for
numerical evolutions. This is in part because these data, just like
standard puncture
data~\cite{Brandt:1997tf,Tichy:2003zg,Tichy:2003qi,Ansorg:2004ds}, do
not contain realistic gravitational waves in the far zone: $h^{{\rm
    TT}\,({\rm NZ})}_{i j}$ does not even vaguely agree with the 2PN
approximation to the waveform amplitude nor with the quadrupole
approximation to the waveform phase for realistic inspiral.

To illustrate this, we restrict to the case of two point sources, and
compute the ``plus'' and ``cross'' polarizations of the near-zone
approximation for $h^{{\rm TT}}_{ij}$:
\bea
h^{({\rm NZ})}_+ &=& h^{{\rm TT}\,({\rm NZ})}_{i j} e_{\theta}^i \, e_{\theta}^j \label{eqn:hplus_NZ},\\
h^{({\rm NZ})}_{\times} &=& h^{{\rm TT}\,({\rm NZ})}_{i j} e_{\theta}^i \, e_{\phi}^j \label{eqn:hcross_NZ}.
\eea
For comparison, the corresponding polarizations of the quadrupole
approximation for the gravitational-wave strain are given by
(paraphrasing Eq.~(3.4) of \cite{Finn:1992xs}):
\bea
h_+\!\!\!&=&\!\!\!\frac{2G\mM}{r} (1\!+\!\cos^2\theta) (\pi G\mM f_{{\rm GW}})^{2/3}\!\cos(\Phi_{{\rm GW}}),\label{eqn:hplus_quad}\\
h_{\times}\!\!\!&=&\!\!\!\frac{4G\mM}{r} \cos\theta (\pi G\mM f_{{\rm GW}})^{2/3}\!\sin(\Phi_{{\rm GW}}) \label{eqn:hcross_quad},
\eea
where $\mM \equiv \nu^{3/5}\,M$ is the ``chirp mass'' of the binary,
given in terms of the total PN mass of the system $M = m_1 + m_2$, and
the symmetric mass ratio $\nu = m_1 m_2 / M^2$.  The angle $\theta$ is
the ``inclination angle of orbital angular momentum to the line of
sight toward the detector''; that is, just the polar angle to the
field point, when the binary moves in the $x$-$y$ plane.  $\Phi_{{\rm GW}}$
and $f_{{\rm GW}}$ are the phase and frequency of the radiation at time $t$,
exactly twice the orbital phase $\Phi(t - r)$ and orbital frequency
$\Omega(t - r)/2\pi$.

\begin{figure}
  \begin{center}
    \includegraphics*[width=3.5in]{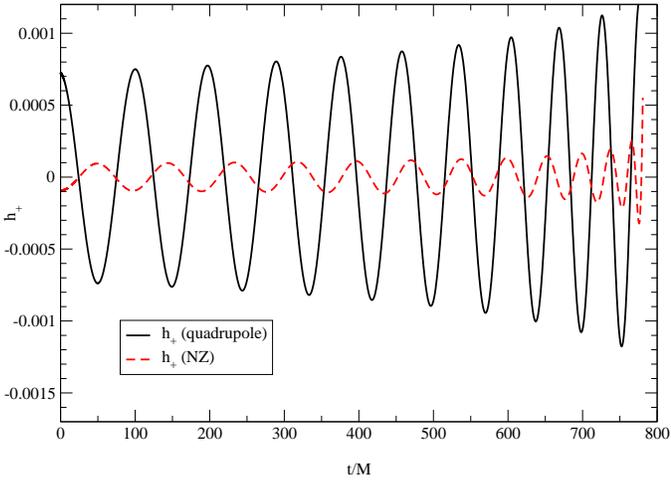}
  \end{center}
  \caption{\label{fig:strain_JS-quad}
  Plus polarization of the quadrupole (black/solid) and near-zone
  (red/dashed) strains observed at field point $r=100M$, $\theta =
  \pi/4$, $\phi = 0$. The binary orbits in the $x$-$y$ plane, with
  initial separation $r_{12} = 10 M$, and a nominal coalescence time
  $t_c \approx 780M$. Both phase and amplitude of $h^{{\rm
  TT}\,(4)}_{ij}$ are very wrong outside the near zone.}
\end{figure}

The lowest-order PN prediction for radiation-reaction effects yields a
simple inspiral of the binary over time, with orbital phasing given by
\bea
\Phi(\tau) = \Phi(t_c) - \frac{1}{\nu} \Theta^{5/8} \label{eqn:phSimpleInspiral}, \\
\Omega(\tau) = \frac{1}{8GM} \Theta^{-3/8} \label{eqn:omSimpleInspiral},
\eea
where $\Theta \equiv \nu \, (t_c - \tau) / 5\,GM$, $M$ and $\nu$ are
given below (\ref{eqn:hcross_quad}), and $t_c$ is a nominal
``coalescence time''.  To evaluate
(\ref{eqn:hplus_NZ}-\ref{eqn:hcross_NZ}), we need the transverse
momentum $p$ corresponding to the desired separation $r_{12}$. The
simplest expression for this is the classical Keplerian relation,
which we give parameterized by $\Omega(\tau)$:
\bea
r_{12} &=& G^{1/3} M (M \Omega)^{-2/3}, \label{eqn:DofOmegaNewtonian}\\ 
p &=& M \nu (G M \Omega)^{1/3}. \label{eqn:PofOmegaNewtonian}
\eea
In Fig. \ref{fig:strain_JS-quad} we compare the plus polarization of
the two waveforms (\ref{eqn:hplus_NZ}) and (\ref{eqn:hplus_quad}) at a
field point $r=100M$, $\theta = \pi/4$, $\phi = 0$, for a binary in
the $x$-$y$ plane, with initial separation $r_{12} = 10M$. The orbital
frequency of the binary is related to the separation $r_{12}$ and
momenta $p$ entering (\ref{eqn:hplus_NZ}) by
(\ref{eqn:DofOmegaNewtonian}-\ref{eqn:PofOmegaNewtonian}). To this
level of approximation, the binary has a nominal PN coalescence time
$t_c \approx 780 M$.  As might have been anticipated, both phase and
amplitude of $h^{{\rm TT}\,(4)}_{ij}$ are wrong outside the near
zone. This means that the data constructed from $h^{{\rm
    TT}\,(4)}_{ij}$ have the wrong wave content, but nevertheless
these data are still accurate up to order $(v/c)^3$ in the far zone.

It is evident from the present-time dependence of (\ref{eqn:JS_htt4})
that it cannot actually contain any of the past history of an
inspiralling binary. We would expect that a correct ``wave-like''
contribution should depend rather on the retarded time of each
contributing point source. It seems evident that the correct behavior
must, in fact, be contained in the as-yet unevaluated parts of
(\ref{eqn:full_htt4}). The requisite evaluation is what we undertake
in the next section.

\section{Completing the Evaluation of $h^{{\rm TT}}_{i j}$}\label{Sec:hTT_calc}

To move forward, we will simplify (\ref{eqn:full_htt4}) and
(\ref{eqn:JS_htt4}) to the case of only two particles.  Then
(\ref{eqn:full_htt4}) reduces to:
\begin{widetext}
\bea
h^{{\rm TT}}_{i j} &=& h^{{\rm TT}\,({\rm NZ})}_{i j} + 16\pi\,G \, \int
\left[ \frac{p_{1 \, i} \, p_{1 \, j}}{m_1} \, e^{i \, \vec{k} \, \cdot
(\vec{x} - \vec{x}_1)} + \frac{p_{2 \, i} \, p_{2 \, j}}{m_2} \, e^{i \,
\vec{k} \, \cdot (\vec{x} - \vec{x}_2)} - \frac{G}{2} \, m_1 \, m_2 \,
\frac{n_{12i} \, n_{12j}}{r_{12}} \, e^{i \, \vec{k} \, \cdot (\vec{x} - \vec{x}_1)} \right. \nonumber \\
&& \left. - \frac{G}{2} \, m_2 \, m_1 \, \frac{n_{21i} \, n_{21j}}{r_{12}} \, e^{i \, \vec{k} \, \cdot (\vec{x} - \vec{x}_2)} \right]^{{\rm TT}}_{\tau} \, \cdot \frac{(\omega/k)^2 \, e^{- i \, \omega \, (t - \tau)}}{k^2 - (\omega + i \, \epsilon)^2} \, \frac{d^3\vec{k} \, d\omega \, d\tau}{(2 \, \pi)^4} + h^{{\rm TT}}_{i j, (div)}\\
&=& h^{{\rm TT}\,({\rm NZ})}_{i j} + H^{{\rm TT}\,1}_{i j}\left[ \frac{\vec{p}_1}{\sqrt{m_1}}\right] + H^{{\rm TT}\,2}_{i
j}\left[ \frac{\vec{p}_2}{\sqrt{m_2}}\right] - H^{{\rm TT}\,1}_{i j}\left[ \sqrt{\frac{G\,m_1 \, m_2}{2 \, r_{12}}} \,
\hat{n}_{12}\right]  - H^{{\rm TT}\,2}_{i j}\left[ \sqrt{\frac{G\,m_1 \, m_2}{2 \, r_{12}}} \, \hat{n}_{12}\right] \nonumber \\
&& + h^{{\rm TT}}_{i j, (div)} \label{eqn:htt4_from_H} ,
\eea
where
\bea
\label{eqn:HTT_gendef}
H^{{\rm TT}\,A}_{i j} [\vec{u}] &\defn& 16\pi\,G \, \int d\tau \, \frac{d^3\vec{k} \, d\omega}{(2 \, \pi)^4} \, [u_i \, u_j]^{{\rm TT}}_{\tau} \, \frac{(\omega/k)^2}{k^2 - (\omega + i \, \epsilon)^2} e^{i \, \vec{k} \cdot (\vec{x} - \vec{x}_A(\tau))} \, e^{- i \, \omega \, (t - \tau)}.
\eea
\end{widetext}
Here, the ``TT projection'' is effected using the operator $P_i^j
\defn \delta_i^j - k_i \, k^j / k^2$. For an arbitrary spatial vector
$\vec{u}$,
\bea
\label{eqn:uTT_mom}
[u_i \, u_j]^{{\rm TT}} &=& u_c \, u_d \, ( P_i^c \, P_j^d - \frac{1}{2} \, P_{i j} \, P^{c d} ) \nonumber\\
&=& u_i \, u_j + \frac{1}{2} \left[ \left( \frac{u_c \, k^c}{k} \right)^2 - u^2 \right] \delta_{i j} \nonumber \\
&& + \frac{1}{2} \left[  \left( \frac{u_c \, k^c}{k} \right)^2 + u^2 \right] \, \frac{k_{i} \, k_{j}}{k^2} \nonumber \\
&& - 2 \, \left( \frac{u_c \, k^c}{k} \right) \, \frac{u_{(i} \, k_{j)}}{k}.
\eea

Details on the evaluation of these terms are presented in Appendix
\ref{sec:app_calc}. After calculation, we write the result as a sum of
terms evaluated at the \emph{present} field-point time $t$, the
\emph{retarded} time $t_A^r$ defined by
\bea
\label{eqn:tret_def}
t - t_A^r - r_A(t_A^r) = 0 , 
\eea
and \emph{integrals} between $t_A^r$ and $t$,
\bea
H^{i \, j}_{{\rm TT}\,A} [\vec{u}] &=& H^{i \, j}_{{\rm TT}\,A} [\vec{u};t] + H^{i \, j}_{{\rm TT}\,A} [\vec{u};t_A^r] \nonumber \\
&& + H^{i \, j}_{{\rm TT}\,A} [\vec{u};t_A^r \rightarrow t],
\eea
where the three parts are given by:
\begin{widetext}
\bea
H^{i \, j}_{{\rm TT}\,A} [\vec{u};t] &=& - \frac{1}{4} \, \frac{G}{r_A(t)} \left\{  \left[ u^2  - 5 \, (\vecdothat{u}{n}_A)^2 \right] \, \delta^{i \, j} + 2 \, u^i \, u^j  + \left[ 3 \, (\vecdothat{u}{n}_A)^2 - 5 \, u^2 \right] \, n_A^i \, n_A^j \right. \nonumber \\
&& \left. + 12 \, (\vecdothat{u}{n}_A) \, u^{(i} \, n_A^{j)} \right\}_{t}, \label{eqn:H_present} \\
H^{i \, j}_{{\rm TT}\,A} [\vec{u};t_A^r] &=& \frac{G}{r_A(t_A^r)} \left\{ \left[ - 2\,u^2 + 2\,(\vecdothat{u}{n}_A)^2 \right] \, \delta^{i \, j} + 4 \, u^i \, u^j + \left[ 2\,u^2 + 2\,(\vecdothat{u}{n}_A)^2 \right] \, n_A^i \, n_A^j  \right. \nonumber \\
&& \left. - 8 \, (\vecdothat{u}{n}_A) \, u^{(i} \, n_A^{j)} \right\}_{t_A^r}, \label{eqn:H_retarded} \\
H^{i \, j}_{{\rm TT}\,A} [\vec{u};t_A^r \rightarrow t] &=& - G \, \int_{t_A^r}^{t} d\tau \, \frac{(t - \tau)}{r_A(\tau)^3} \, \left\{ \left[ - 5 \, u^2  + 9 \, (\vecdothat{u}{n}_A)^2  \right] \, \delta^{i \, j} + 6 \, u^i \, u^j - 12 \, (\vecdothat{u}{n}_A)\, u^{(i} \, n_A^{j)}  \right. \nonumber\\
&& \left. + \left[ 9 \, u^2 - 15 \, (\vecdothat{u}{n}_A)^2 \right] \, n_A^i \, n_A^j \right\} \nonumber\\
&& - G \, \int_{t_A^r}^{t} d\tau \, \frac{(t - \tau)^3}{r_A(\tau)^5} \, \left\{ \left[ u^2 - 5 \, (\vecdothat{u}{n}_A)^2 \right] \, \delta^{i \, j} + 2 \, u^i \, u^j - 20 \, (\vecdothat{u}{n}_A) \, u^{(i} \, n_A^{j)} \right.\nonumber \\
&& \left. + \left[ - 5 \, u^2 + 35 \, (\vecdothat{u}{n}_A)^2 \right] \, n_A^i \, n_A^j  \right\}. \label{eqn:H_interval}
\eea
\end{widetext}

\begin{figure}
  \begin{center}
    \includegraphics*[width=3.5in]{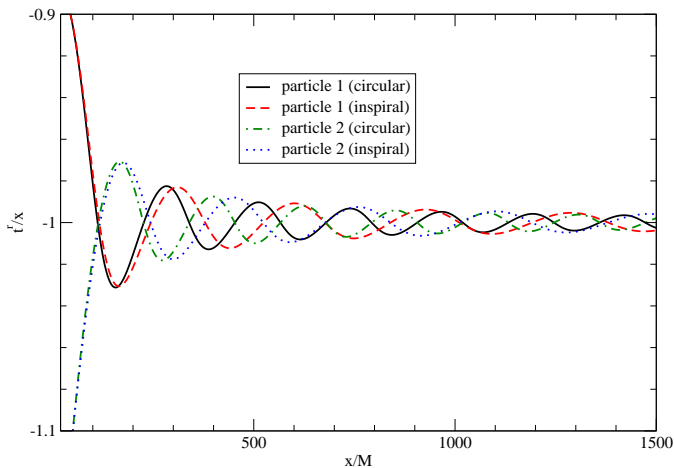}
  \end{center}
  \caption{\label{fig:tret_orbit_comp}
    Retarded times for particles 1 and 2, as measured by observers
    along the $x$ axis at the initial time $t = 0$, for the binary of
    Fig. \ref{fig:strain_JS-quad}. To highlight the oscillatory effect
    of the finite-radius orbit on $t^r$, we first divide by the average
    field distance $r$.}
\end{figure}

In Fig. \ref{fig:tret_orbit_comp}, we show the retarded times
calculated for each particle, as measured at points along the $x$
axis, for the same orbit as in Fig. \ref{fig:strain_JS-quad}. We also
show the corresponding retarded times for a binary in an exactly
circular orbit. Since the small-scale oscillatory effect of the finite
orbital radius would be lost by the overall linear trend, we have
multiplied by the orbital radius.

\subsection{Reconciling with Jaranowski \& Sch\"{a}fer's $h^{{\rm TT}\,(4)}_{ij}$}

From the derivation above it is clear that $h^{{\rm TT}}_{ij}$
includes retardation effects, so it will not depend solely on the
present time. We might even expect that all ``present-time''
contributions should vanish individually, or should cancel out.  It
can be seen easily from (\ref{eqn:H_present}) that the ``$t$'' part of
the second and third terms of Eq.~(\ref{eqn:htt4_from_H}) exactly
cancel out the ``kinetic'' part (first line) of
Eq.~(\ref{eqn:JS_htt4}).  Thus, we can simply remove that line in
Eq.~(\ref{eqn:JS_htt4}), and use the ``$t^r$'' part instead. One may
similarly inquire whether the ``$t$'' parts of the fourth and fifth
terms of Eq.~(\ref{eqn:htt4_from_H}) above,
\bea
h^{{\rm TT}\,(pot,now)}_{i j} &\equiv& - H^{{\rm TT}\,1}_{i j}\left[ \sqrt{\frac{G\,m_1 \, m_2}{2 \, r_{12}}} \,
\hat{n}_{12};t\right] \nonumber\\
&&  - H^{{\rm TT}\,2}_{i j}\left[ \sqrt{\frac{G\,m_1 \, m_2}{2 \, r_{12}}} \, \hat{n}_{12};t\right],
\eea
also cancel the remaining, ``potential'' parts of
Eq.~(\ref{eqn:JS_htt4}). The answer is ``not completely''; expanding
in powers of $1/r$, we find:
\begin{widetext}
\bea
h^{{\rm TT}\,(pot,4)}_{ij} + h^{{\rm TT}\,(pot,now)}_{i j} &=& \frac{G^2\,m_1\,m_2\,r_{12}}{16\,r^3} \left\{ (3 + 14\,W^2 - 25\,W^4) \, \delta_{i \, j} - 4\,(1 + 5\,W^2) \, n_{12i} \, n_{12j} \right. \nonumber \\
&& \left. - 5\,(1 + 6\,W^2 - 7\,W^4) \, n_{1i} \, n_{1j} + 2\,W\,(7 + 9\,W^2) \, \left( n_{12i} \, n_{1j} + n_{12j} \, n_{1i} \right)\right\} + O(1/r^4),
\eea
\end{widetext}
where $W \equiv \sin\theta \, \cos(\phi - \Phi(t))$, and $\Phi(t)$ is
the orbital phase of particle 1 at the present time $t$. That is, the
``new'' contribution cancels the $1/r$ and $1/r^2$ pieces of $h^{{\rm
    TT}\,(4)}_{ij}$ entirely.  In the far zone the result is thus
smaller than the $h^{{\rm TT}}_{ij, (div)}$ term which we are ignoring
everywhere, since it is small both in the near and the far
zone~\cite{Schafer:1986rd}.

We note here two general properties of the contributions to the full
$h^{{\rm TT}}_{i j}$.
\begin{enumerate}
\item In the near zone $h^{{\rm TT}\,(4)}_{ij}$ is the dominant term
  since all other terms arise from $(\Box_{ret}^{-1} -\Delta^{-1})
  s_{kl}$.  Thus all other terms must cancel within the accuracy of
  the near-zone approximation.
\item $h^{{\rm TT}\,(4)}_{ij}$ is wrong far from the sources;
  thus, the new corrections should ``cancel'' $h^{{\rm TT}\,(4)}_{ij}$
  entirely, far from sources. Note, however, that while $h_{ij} = -
  \Box_{ret}^{-1} s_{kl}$ depends only on retarded time, its
  TT-projection $h^{{\rm TT}}_{ij} = \delta^{{\rm TT}\, kl}_{ij}
  h_{kl}$ has a more complicated causal structure; E.g. the finite
  time integral comes from applying the TT-projection.  [Proof: Even
  if we had a source given exactly by $\bar{s}_{kl}$, $h^{{\rm
      TT}\,(4)}_{ij}$ would depend only the present time, $h_{ij}$
  would depend only on retarded time, and $h^{{\rm TT}}_{ij}$ would
  (as we have computed) contain a finite time integral term.]
\end{enumerate}
Additionally, the full $h^{{\rm TT}}_{ij}$ agrees well with quadrupole
predictions, which we demonstrate in Section \ref{Sec:numerics}.

\section{Numerical Results and Invariants}\label{Sec:numerics}

\subsection{Phasing and Post-Keplerian Relations}

It has been known for some time (see for example \cite{Cutler:1993tc})
that gravitational wave phase plays an even more important part in
source identification than does wave amplitude.  In PN work,
phase and amplitude are estimated somewhat separately; the
amplitude requires knowledge of the time-dependent multipoles, used in
developing the the full metric, while the phase can be relatively
simply approximated from the orbital equations of motion, taking into
account the gravitational wave flux at infinity to evolve the orbital
parameters \cite{Tichy:1999pv}.

The quadrupole waveform introduced for the comparison in Fig.\
\ref{fig:strain_JS-quad} had an amplitude accurate to $O(v/c)^4$ and
the simplest available time evolution for the phase.  Waveform phase
is a direct consequence of orbital phase.  To lowest order, we could
have assumed a binary moving in a circular orbit (of zero
eccentricity) since, up to 2PN order, we can have circular orbits,
where the linear momentum, $p$, of each particle is related to the
separation $r_{12}$ by, say, Eq.~(24) of \cite{Tichy:2002ec}.
Nevertheless, circular orbits are physically unrealistic -- since
radiation reaction will lead to inspiral and merger of the particles
-- and Eqs.~(\ref{eqn:phSimpleInspiral}-\ref{eqn:omSimpleInspiral})
already include leading-order radiation-reaction effects.  Moreover,
the phase errors that would accrue from using purely circular orbits
would be larger, the further from the sources we tried to compute
them.

The calculations of section \ref{Sec:hTT_calc} lead to waveform
amplitudes that are accurate at $O(v/c)^4$ everywhere.  However, we
desire that our initial-data wave content already encode the phase as
accurately as possible.  Highly accurate \emph{phase} for our initial
data (via $h^{{\rm TT}}$), and hence in the leading edge of the
waveforms we would extract from numerical evolution, is critical for
parameter estimation following a detection.

For demonstrative purposes, in this section, we will restrict
ourselves to the simplest phasing relations consistent with
radiation-reaction inspiral as given by Eqs.\
(\ref{eqn:phSimpleInspiral}-\ref{eqn:omSimpleInspiral}), while using
higher-order PN expressions than Eqs.\ (\ref{eqn:DofOmegaNewtonian}
-\ref{eqn:PofOmegaNewtonian}) for relating the orbit to the phase.
For example, from \cite{SchaferWex_1993}, we have found to second PN
(beyond leading) order:
\bea
\frac{r_{12}(\Omega)}{G M} &=& (G M \Omega)^{-2/3} - \frac{(3 - \nu)}{3} \nonumber \\
 && - \frac{(18 - 81 \nu - 8 \nu^2)}{72} (G M \Omega)^{2/3}, \label{eqn:DofOmega}\\
\frac{p(\Omega)}{M \nu} &=& (G M \Omega)^{1/3} + \frac{(15 - \nu)}{6} (G M \Omega) \nonumber \\
&& + \frac{(441 - 324 \nu - \nu^2)}{72} (G M \Omega)^{5/3}, \label{eqn:PofOmega}
\eea
and we note that higher-order equivalents of these can be computed
from \cite{Memmesheimer:2004cv}.

In the numerical construction of initial data, the primary input is
the coordinate separation of the holes.  In placing the punctures on
the numerical grid, the separation must be maintained exactly.  To
ensure this, we invert Eq.~(\ref{eqn:DofOmega}) to obtain the exact
$\Omega_r$ corresponding to our desired $r_{12}$.  Then we use
Eq.~(\ref{eqn:omSimpleInspiral}) with $t = 0$ to find the coalescence
time $t_c$ that yields this $\Omega_r$.  Once we have obtained $t_c$,
we then find the orbital phase $\Phi$ and frequency $\Omega$ at any
source time $\tau$ directly from
Eqs.~(\ref{eqn:phSimpleInspiral}-\ref{eqn:omSimpleInspiral}), and the
corresponding separation $r_{12}$ and momentum $p$ from Eqs.
(\ref{eqn:DofOmega}-\ref{eqn:PofOmega}), or their higher-order
equivalents.

\begin{figure}
  \begin{center}
    \includegraphics*[width=3.5in]{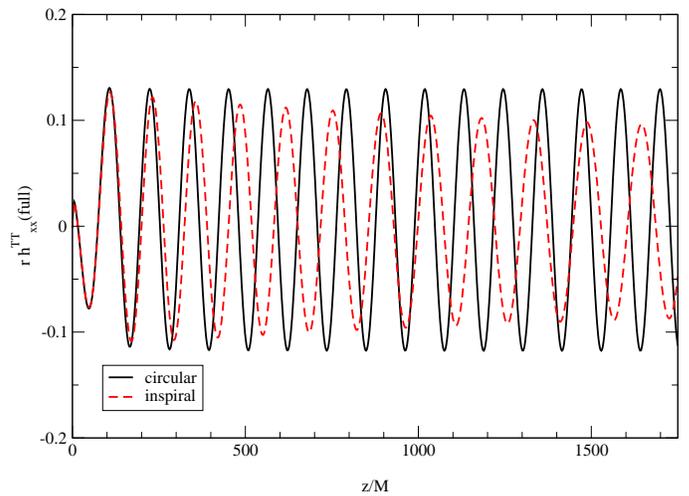}
  \end{center}
  \caption{\label{fig:hTT_circ_insp_comp}
  The $xx$ component of the full $h^{{\rm TT}}_{ij}$ for a binary with
  initial separation $r_{12}=10M$ in a circular (black/solid) or
  inspiralling (red/dashed) orbit. Both fields have been rescaled by
  the observer radius $r = z$ to compensate for the leading $1/r$ fall-off.
  The orbital configuration is the
  same as for Fig. \ref{fig:strain_JS-quad}, apart from the Keplerian
  relations, where we have used the higher-order relations
  (\ref{eqn:DofOmega}-\ref{eqn:PofOmega}), yielding $t_c \sim 1100 M$. Note the frequency
  broadening at more distant field points.}
\end{figure}

In Fig. \ref{fig:hTT_circ_insp_comp}, we show a representative
component of the retarded-time part of $h^{{\rm TT}}_{ij}$ for both
circular and leading-order inspiral orbits. For both orbits, we use
the extended Keplerian relations (\ref{eqn:DofOmega}) and
(\ref{eqn:PofOmega}); otherwise the orbital configuration is that of
Fig. \ref{fig:strain_JS-quad}. The coalescence time is now $t_c \sim
1100 M$. We can see that the cumulative wavelength error of the
circular-orbit assumption becomes very large at large distances from
the sources.
This demonstrates that using inspiral orbits instead of circular
orbits will significantly enhance the phase accuracy of the initial
data, even though circular orbits are in principle sufficient when we
include terms only up to $O(v/c)^4$ as done in this work. From now on
we use only inspiral orbits.

\begin{figure}
  \begin{center}
    \includegraphics*[width=3.5in]{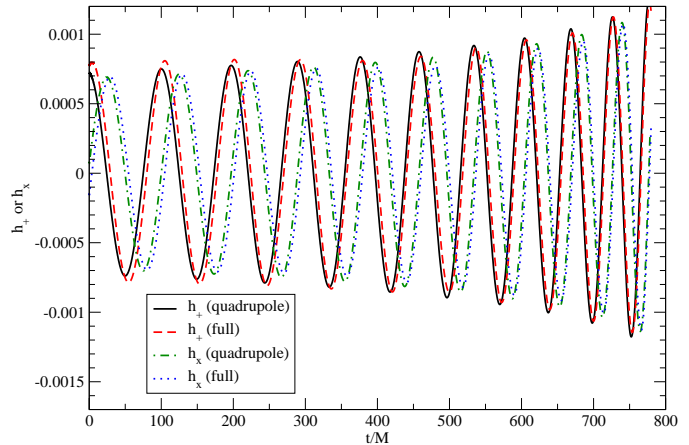}
  \end{center}
  \caption{\label{fig:strain_quad-full}
    Plus and cross polarizations of the strain observed at field point
    $r=100M$, $\theta = \pi/4$, $\phi = 0$.  Both the
    quadrupole-approximation waveform (black/solid and green/dot-dashed)
    and the full (red/dashed and blue/dotted)
    waveforms coming from $h^{{\rm TT}}_{ij}$ are shown. The orbital
    configuration is the same as for Fig. \ref{fig:strain_JS-quad}.
}
\end{figure}

\begin{figure}
  \begin{center}
    \includegraphics*[width=3.5in]{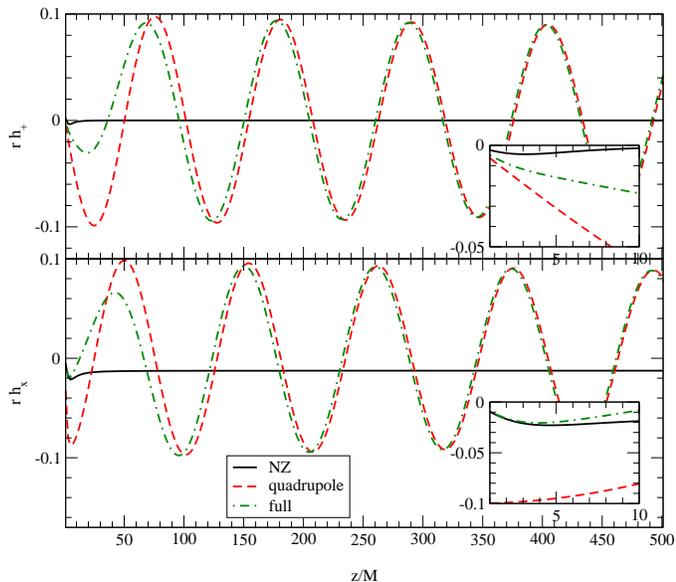}
  \end{center}
  \caption{\label{fig:strain_quad-full-NZ_space}
    Plus and cross polarizations of the strain observed at $t=0$ along
    the $z$ axis. We show the near-zone (solid/black), the quadrupole
    (dashed/red) and full (dot-dashed/green) waveforms. All waveforms
    have been rescaled by the observer radius $r = z$ to compensate
    for the leading $1/r$ fall-off. The orbital configuration is the same
    as for Fig. \ref{fig:strain_JS-quad}.
}
\end{figure}

Next, we compare our full waveform $h^{{\rm TT}}_{ij}$ (expressed as
the combinations $h_+$ and $h_{\times}$) at an intermediate-field
position ($r=100M$, $\theta = \pi/4$, $\phi = 0$) to the lowest-order
quadrupole result.  In Fig. \ref{fig:strain_quad-full}, the orbital
configuration is the same as for Fig. \ref{fig:strain_JS-quad}.
As one can see, both the $+$ and $\times$ polarizations of our
$h^{{\rm TT}}_{ij}$ agree very well with quadrupole results, as they
should.  We demonstrate the near- and intermediate-zone behavior of
the new data on the initial spatial slice in
Fig. \ref{fig:strain_quad-full-NZ_space}. The quadrupole and full
solutions agree very well outside $\sim 100 M$. However, the full
solution's phase and amplitude approach the NZ solution closer to the
sources.

\subsection{Numerical Implementation}

After having confirmed that we have a PN three-metric $g_{ij}$ that is
accurate up to errors of order $O(v/c)^{5}$, and that correctly
approaches the quadrupole limit outside the near zone, we are now
ready to construct initial data for numerical evolutions.  In order to
do so, we need the intrinsic curvature $K_{ij}$, which can be computed
as in Tichy \etal \cite{Tichy:2002ec} from the conjugate momentum. The
difference is that here we use the full $\dot{h}^{{\rm TT}}_{ij}$
instead of the near-zone approximation $\dot{h}^{{\rm TT}\,(4)}_{ij}$
to obtain the conjugate momentum~\cite{Schafer:1986rd}. The result is
\bea
\label{Kij_up}
K^{ij} &=& -\psi_{PN}^{-10} \left[ \tilde{\pi}^{ij}_{(3)} 
                + \frac{1}{2}\dot{h}_{ij}^{{\rm TT}}
                + (\phi_{(2)}\tilde{\pi}^{ij}_{(3)})^{{\rm TT}} 
                \right] 
\nonumber \\
&& + O(v/c)^{6} ,
\eea
where the error term comes from neglecting terms like $h^{{\rm
    TT}}_{ij, (div)}$ at $O(v/c)^{5}$ in ${h}^{{\rm TT}}_{ij}$, and
where $\psi_{PN}$, $\tilde{\pi}^{ij}_{(3)}$ and $\phi_{(2)}$ can be
found in Tichy \etal \cite{Tichy:2002ec}.  An additional difference is
that the time derivative of $h^{{\rm TT}}_{ij}$ is evaluated
numerically in this work.  Note that the results for $g_{ij}$ are
accurate up to $O(v/c)^{4}$, while the results for $K_{ij}$ are
accurate up $O(v/c)^{5}$, because $K_{ij}$ contains an additional time
derivative~\cite{Tichy:2002ec,Yunes:2005nn,Yunes:2006iw}.

\begin{figure}
  \begin{center}
    \includegraphics*[width=3.5in]{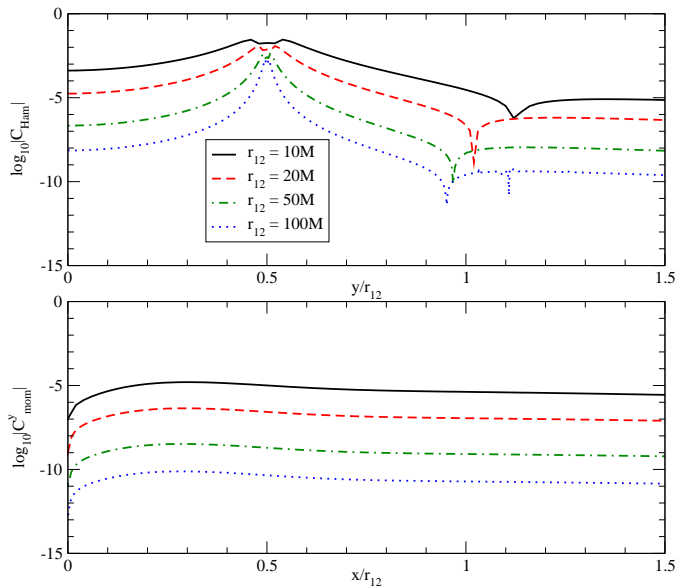}
  \end{center}
  \caption{
  Upper panel: Hamiltonian constraint violation along the $y$ axis of
  our new data in the near zone, as a function of binary separation
  $r_{12}$. Lower panel: Momentum constraint ($y$-component) violation
  of the same data along the $x$ axis. The orbital configuration is
  that of Fig. \ref{fig:hTT_circ_insp_comp}. Distances have been scaled
  relative to $r_{12}$, so that the punctures are initially at
  $y/r_{12} = \pm 0.5$.
}
\label{fig:Cham_Y_Cmom_X_2.5PN}
\end{figure}

Next we show the violations of the Hamiltonian and momentum
constraints computed from $g_{ij}$ and $K_{ij}$, as functions of the
binary separation $r_{12}$. As we can see in both panels of
Fig.~\ref{fig:Cham_Y_Cmom_X_2.5PN}, the constraints become smaller for
larger separations, because the post-Newtonian approximation gets
better.  Note that, as in~\cite{Tichy:2002ec}, the constraint
violation remains finite everywhere, and is largest near each black
hole.

\subsection{Curvature Invariants and Asymptotic Flatness}

In analysis of both initial and evolved data, it is often instructive
to investigate the behavior of scalar curvature invariants, as these
give some idea of the far-field properties of our solution. We expect,
for an asymptotically flat space-time, that in the far field, the
speciality index $\mS \equiv 27 \mJ^2 / \mI^3$ will be close to unity.
This can be seen from the following arguments.  Let us choose a tetrad
such that the Weyl tensor components $\psi_1$ and $\psi_3$ are both
zero. Further, we assume that in the far field $\psi_0$ and $\psi_4$
are both perturbations of order $\epsilon$ off a Kerr background.
Then
\begin{equation}
\mS \approx 1 -  3\frac{\psi_0 \psi_4}{\psi_2^2} + O(\epsilon^3) ,
\end{equation}
which is indeed close to one.  Note however, that this argument only
works if the components of the Weyl tensor obey the peeling theorem,
such that $\psi_2 \sim O(r^{-3})$, $\psi_0 \sim O(r^{-5})$ and $\psi_4
\sim O(r^{-1})$.  In particular, if $\psi_0$ falls off more slowly
than $O(r^{-5})$, $S$ will grow for large $r$. Now observe that
$\psi_0 \sim O(r^{-5}) \sim M^3/r^5$ is formally of $O(v/c)^6$. Thus,
in order to see the expected behavior of $\mS \approx 1$ in the
far-field we need to go to $O(v/c)^6$. If we only go to $O(v/c)^4$ (as
done in this work) $\psi_0$ consists of uncontrolled remainders only,
which should in principle be dropped. When we numerically compute $\mS
$ we find that for our data, $\mS$ deviates further and further from
unity for large distances from the binary. This reflects the fact that
the so-called ``incoming'' Weyl scalar $\psi_0$ only falls off as
$1/r^3$, due to uncontrolled remainders at $O(v/c)^6$, which arise
from a mixing of the background with the TT waveform.

\section{Discussion and Future Work}\label{Sec:discuss}

Exploring and validating PN inspiral waveforms is crucially important
for gravitational-wave detection and for our theoretical understanding
of black-hole binaries.  Our goal has been to provide a step forward
in this understanding by building a direct interface between the PN
approach and numerical evolution, along the lines initially outlined
in Ref. \cite{Tichy:2002ec}.  In this paper we have essentially
completed the calculation of the transverse-traceless part of the
ADM-TT metric to $O(v/c)^4$ provided in \cite{Tichy:2002ec}, yielding
data that, on the initial Cauchy slice, will describe the space-time
into the far-field.  We have incorporated this formulation into a
numerical initial-data routine adapted to the ``puncture'' topology
that has been so successful recently, and have explored these data's
numerical properties on the initial slice.

Our next step is to evolve these data with moving punctures, and
investigate how the explicit incorporation of post-Newtonian waveforms
in the initial data affects both the ensuing slow binary inspiral of
the sources and the release of radiation from the system.  We note
especially that our data are non-conformally flat beyond $O(v/c)^3$.
We expect our data to incorporate smaller unphysical initial
distortions in the black holes than is possible with conformal
flatness, and hence less spurious gravitational radiation during the
numerical evolution.  We see this as a very positive step toward
providing further validation of numerical relativity results for
multiple orbit simulations, since it permits comparison with PN
results where they are expected to be reliable. Our initial data will
also allow us to fully evaluate the validity of PN results for merging
binaries by enabling comparison with the most accurate numerical
relativity results.

We expect that further development of these data will certainly
involve the use of more accurate orbital phasing information than the
leading order given by
Eqs.~(\ref{eqn:phSimpleInspiral}-\ref{eqn:omSimpleInspiral}).  This
information is available in radiative coordinates (see,
e.g. Eq.~(6.29) of \cite{Blanchet:1996wx}) appropriate for far-field
evaluation of the gravitational radiative modes; it may be possible to
produce them in ADM-TT coordinates via a contact transformation, or by
direct calculation (see, e.g.\ \cite{Damour:2004bz}).  For initial
separations similar to the fiducial test case of this paper,
$r_{12}\!=\!10M$, the order necessary for clean matching of the
initial wave content with the new radiation generated in evolution
should not be particularly high \cite{Baker:2006ha}. As noted, the
Keplerian relations Eqs.~(\ref{eqn:DofOmega}-\ref{eqn:PofOmega}) can
easily be extended to higher PN order.

The data presented already allow for arbitrary initial mass ratios
$\nu$; this introduces the possibility of significant gravitational
radiation in odd-$l$ multipoles, together with associated phenomena,
such as in-plane recoil ``kicks''.  An interesting future development
of these data will be the inclusion of spin angular momenta on the
pre-merger holes. This will open our initial-data prescription to
describing an even richer spectrum of binary radiation.

\acknowledgments

We would like to thank L. Blanchet and G. Sch\"afer for generous
assistance and helpful discussion.

M.C., B.K.~and B.W.~gratefully acknowledge the support of the NASA
Center for Gravitational Wave Astronomy (NAG5-13396).  M.C.~and
B.K.~also acknowledge the NSF for financial support under grants
PHY-0354867 and PHY-0722315. B.K.~also acknowledges support from the
NASA Postdoctoral Program at the Oak Ridge Associated Universities.
The work of W.T.~was supported by NSF grant PHY-0555644.  W.T.~also
acknowledges partial support from the NCSA under Grant PHY-060040T.
The work of B.W.~was also supported by NSF grants PHY-0245024 and
PHY-0555484.

\appendix

\section{Details of Integral Calculation}
\label{sec:app_calc}

Here we present some more details of the calculations that lead to the
three contributions to Eq.~(\ref{eqn:HTT_gendef}): Eqs.
(\ref{eqn:H_present}-\ref{eqn:H_interval}). Inserting Eq.
(\ref{eqn:uTT_mom}) in the general integral (\ref{eqn:HTT_gendef}), we
can write $H^{i j}_{{\rm TT}\,A} [\vec{u}]$ as a combination of scalar
and tensor terms:
\bea
H^{{\rm TT}\,A}_{i j} [\vec{u}] &=& 16\pi\,G\int d\tau \left\{ \left[ u_i \, u_j - \frac{u^2}{2} \, \delta_{i j} \right]_{\tau} \, I_A \right. \nonumber \\
& & \left. + \left[ \frac{u^2}{2} \right]_{\tau} \, I_{i j \, A} + \left[ \frac{u_c \, u_d}{2} \right]_{\tau} \, I^{c d}_A \, \delta_{i j} \right. \nonumber \\
& & \left. - \left[ 2 \, u_c \, u_{(i} \right]_{\tau} \, I_{j)_A}^c + \left[ \frac{u_c \, u_d}{2} \right]_{\tau} \, I_{i j \; A}^{\;\;\; c d} \right\}, \label{eqn:HTT_Idecomp}
\eea
where the ``$I$'' integrals are defined as:
\bea
I_A &\equiv& \int \frac{d^3\vec{k} \, d\omega}{(2 \, \pi)^4}\frac{(\omega/k)^2 \, e^{i \, k \, r_A \, \cth  - i \, \omega \, T}}{k^2 - (\omega + i \, \epsilon)^2} , \label{eqn:I0_def}\\
I^{i \, j}_A &\equiv& \int \frac{d^3\vec{k} \, d\omega}{(2 \, \pi)^4}\frac{k^i \, k^j}{k^2} \nonumber \\
&& \times \frac{(\omega/k)^2 \, e^{i \, k \, r_A \, \cth  - i \, \omega \, T}}{k^2 - (\omega + i \, \epsilon)^2}, \label{eqn:I2_def}\\
I^{i \, j \, c \, d}_A &\equiv& \int \frac{d^3\vec{k} \, d\omega}{(2 \, \pi)^4} \frac{k^i \, k^j \, k^c \, k^d}{k^4} \nonumber \\
&& \times \frac{(\omega/k)^2 \, e^{i \, k \, r_A \, \cth  - i \, \omega \, T}}{k^2 - (\omega + i \, \epsilon)^2}. \label{eqn:I4_def}
\eea
Here $T \equiv t - \tau$, and $\vec{r}_A \equiv \vec{x} -
\vec{x}_A$. We have also taken our integration coordinates such that
$\vec{r}_A$ lies in the $z$ direction, so that the dummy momentum
vector $\vec{k}$ satisfies
\bea
\vecdotvec{k}{r}_A &=& k \, r_A \, \cth,\\
d^3\vec{k} &=& k^2 \, dk \, \sth \, d\theta \, d\phi.
\eea

Define the unit orthogonal vectors $\hat{n}_A \equiv ( 0, 0, 1) \; \;
, \; \; \hat{\ell} \equiv ( \cph, \sph, 0)$. Then we can write
\[
\vec{k} = k \, \cth \, \hat{n}_A + k \, \sth \, \hat{\ell} \;\;
\Rightarrow \vecdotvec{k}{r}_A = r_A \, \vec{k} \cdot \hat{n}_A.
\]

We can also define a projector tensor onto $\hat{\ell}$:
\bean
Q^{a \, b} &\equiv& \delta^{a \, b} - n^a \, n^b \Rightarrow Q^a_{\;\; b} = \delta^a_{\;\; b} - n^a \, n_b\\
\Rightarrow Q^a_{\;\; c} \, Q^c_{\;\; b} &=& Q^a_{\;\; b} \;\; , \;\; Q^a_{\;\; b} \, n^b = 0 \;\; , \;\; Q^a_{\;\; b} \, \ell^b = \ell^a.
\eean

\subsection{Angular integration}

We will neglect the $A$ subscript for now, until it becomes relevant
again. To calculate the integrals (\ref{eqn:I0_def}-\ref{eqn:I4_def}),
we begin with the $\phi$ integration. The only $\phi$ dependence comes
from the $\vec{\ell}$ parts of the $\vec{k}$ terms. It can be seen
from elementary trigonometric integrals that:
\bean
\int d\phi \, \ell^a &=& \int d\phi \, \ell^a \, \ell^b \, \ell^c = 0,\;\; \int d\phi \, \ell^a \, \ell^b = \pi \, Q^{a \, b},\\
\int d\phi \, \ell^a \, \ell^b \, \ell^c \, \ell^d &=& \frac{\pi}{4} \left( Q^{a \, b} \, Q^{c \, d} + Q^{a \, c} \, Q^{b \, d} + Q^{a \, d} \, Q^{b \, c} \right).
\eean

We use these to calculate the $\phi$ integrals for $I^{a \, b}_A$ and
$I^{a \, b \, c \, d}_A$. Define $w \equiv \cth$. Then
\bean
\int d\phi \, 1 &=& 2 \, \pi,\\
\int d\phi \frac{k^a \, k^b}{k^2} &=& 2 \, \pi \, w^2 \, n^a \, n^b + \pi \, (1 - w^2) \, Q^{a \, b},\\
\int d\phi \frac{k^a \, k^b \, k^c \, k^d}{k^4} &=& 2 \, \pi \, w^4 \, n^a \, n^b \, n^c \, n^d \\
 && + 6 \, \pi \, w^2 \, (1 - w^2) \, Q^{(a \, b} \, n^c \, n^{d)} \\
 && + \frac{3 \, \pi}{4} \, (1 - w^2)^2 \, Q^{(a \, b} \, Q^{c \, d)}.
\eean

So the next integrals will differ in their $\theta$ dependence,
contained in the powers of $w$ above. The $\theta$ integrals will
contain the following basic types:
\bea
g_0(a) &\equiv& \int_{-1}^{+1} dw \, e^{a \, w} = 2 \, \frac{\sinh a}{a},\\
g_2(a) &\equiv& \int_{-1}^{+1} dw \, w^2 \, e^{a \, w} = 2 \, \frac{\sinh a}{a} - 4 \, \frac{\cosh a}{a^2} \nonumber \\
&& + 4 \, \frac{\sinh a}{a^3}\\
g_4(a) &\equiv& \int_{-1}^{+1} dw \, w^4 \, e^{a \, w} = 2 \, \frac{\sinh a}{a} - 8 \, \frac{\cosh a}{a^2} \nonumber \\
&& + 24 \, \frac{\sinh a}{a^3} - 48 \, \frac{\cosh a}{a^4} + 48 \, \frac{\sinh a}{a^5}.
\eea
Now $I^{a \, b}$ and $I^{a \, b \, c \, d}$ can be written as the
linear combinations:
\bea
I^{a \, b} &=& \frac{1}{2} \, \left[ Q^{a \, b} \, I \right]_{\tau} + \left[ (n^a \, n^b - \frac{1}{2} Q^{a \, b}) K \right]_{\tau},\\
I^{a \, b \, c \, d} &=& \left[ \left( n^a \, n^b \, n^c \, n^d - 3 \, Q^{(a \, b} \, n^c \, n^{d)} + \frac{3}{8} \, Q^{(a \, b} \,Q^{c \, d)}\right) L \right]_{\tau} \nonumber \\
&& + \left[ \left(3 \, Q^{(a \, b} \, n^c \, n^{d)} - \frac{3}{4} \, Q^{(a \, b} \,Q^{c \, d)}\right) K \right]_{\tau} \nonumber \\
&& + \frac{3}{8} \, \left[ Q^{(a \, b} \,Q^{c \, d)} \, I \right]_{\tau}.
\eea

$I$ here can be expressed in terms of $g_0(a)$ above:
\bea
I &\equiv& \int \frac{d\omega}{(2 \, \pi)^3} \, \int \frac{d^3\vec{k}}{2 \, \pi} \, \frac{(\omega/k)^2}{k^2 - (\omega + i \, \epsilon)^2} \, e^{i \, k \, r \, \cth - i \, \omega \, T} \nonumber\\
 &=& \int \frac{d\omega}{(2 \, \pi)^3}  \, \omega^2 \, e^{- i \, \omega \, T} \, \int_{-\infty}^{\infty} dk \, \frac{1/2}{k^2 - (\omega + i \, \epsilon)^2} \, g_0(i \, k \, r) \nonumber \\
 &=& \int \frac{d\omega}{(2 \, \pi)^3}  \, \omega^2 \, e^{- i \, \omega \, T} \, J_0. \label{eqn:I_def}
\eea
The $1/2$ factor is because we moved to integrating $k$ over the whole
real line instead of the positive half-line (this is permissible as
$g_n(a)$ is an even function of $a$). $K$ and $L$ are defined
analogously to $I$, but with extra even powers of $\cth = w$:
\bea
K &\equiv& \int \frac{d\omega}{(2 \, \pi)^3} \, \int \frac{d^3\vec{k}}{2 \, \pi} \, \frac{(\omega/k)^2}{k^2 - (\omega + i \, \epsilon)^2} \, e^{i \, k \, r \, \cth - i \, \omega \, T} \ccth \nonumber\\
 &=& \int \frac{d\omega}{(2 \, \pi)^3}  \, \omega^2 \, e^{- i \, \omega \, T}\, \int_{-\infty}^{\infty} dk \, \frac{1/2}{k^2 - (\omega + i \, \epsilon)^2} \, g_2(i \, k \, r) \nonumber\\
 &=& \int \frac{d\omega}{(2 \, \pi)^3}  \, \omega^2 \, e^{- i \, \omega \, T} \, J_2, \label{eqn:K_def}\\
L &\equiv& \int \frac{d\omega}{(2 \, \pi)^3} \, \int \frac{d^3\vec{k}}{2 \, \pi} \, \frac{(\omega/k)^2}{k^2 - (\omega + i \, \epsilon)^2} \, e^{i \, k \, r \, \cth - i \, \omega \, T} \cos^4\theta \nonumber\\
 &=& \int \frac{d\omega}{(2 \, \pi)^3} \, \omega^2 \, e^{- i \, \omega \, T} \, \int_{-\infty}^{\infty} dk \, \frac{1/2}{k^2 - (\omega + i \, \epsilon)^2} \, g_4(i \, k \, r) \nonumber\\
 &=& \int \frac{d\omega}{(2 \, \pi)^3}  \, \omega^2 \, e^{- i \, \omega \, T} \, J_4. \label{eqn:L_def}
\eea

\subsection{Momentum integration}

\begin{figure*}
  \begin{center}
    \includegraphics*{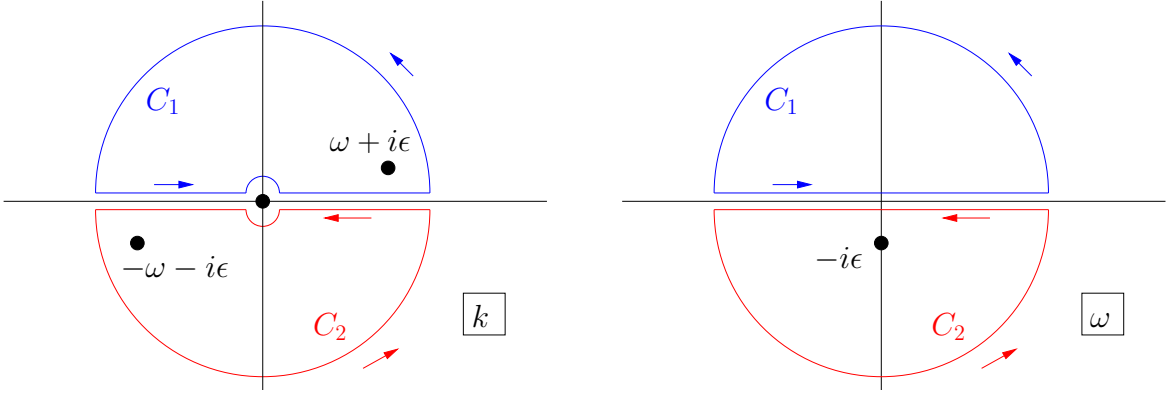}
  \end{center}
  \caption{Contours needed to complete integration over $k$ (left) and $\omega$ (right).}
  \label{fig:contours}
\end{figure*}

Now we address the $k$ integrals, defined as:
\[
J_n \equiv \int_{-\infty}^{\infty} dk \, f_n(k) = \int_{-\infty}^{\infty} dk \, f^+_n(k) + \int_{-\infty}^{\infty} dk \, f^-_n(k),
\]
where we collect the positive exponents in the $g_n$ in the integrand
of $f^+_n(k)$, and the negative exponents in $f^-_n(k)$:
\[
 f^+_n(k) \equiv \frac{g^+_n(i \, k \, r) / 2}{k^2 - (\omega + i \, \epsilon)^2} \;\; , \;\; f^-_n(k) \equiv \frac{g^-_n(i \, k \, r) / 2}{k^2 - (\omega + i \, \epsilon)^2}.
 \]
 We calculate this as the sum of contour integrals of the ``plus'' and
 ``minus'' integrands (necessary, as the opposite signs require
 different contours). Each of these has poles at $k = 0$, $k = k_+
 \equiv \omega + i \, \epsilon$, and $k = k_- \equiv - \omega - i \,
 \epsilon$ (the first of these is from the $g_n$). We integrate the
 ``plus'' integrands anticlockwise around the contour $C_1$,
 and the ``minus'' integrands anticlockwise around the contour $C_2$
 (see Fig. \ref{fig:contours}); taking the limit $|k|
 \rightarrow \infty$, the contribution from the curved segments
 vanishes, and the residue theorem gives us:
\bea
J_n &=& 2 \, \pi \, i \, \mbox{Res}[f^+_n, k_+] - 2 \, \pi \, i \, \mbox{Res}[f^-_n, k_-] \nonumber \\
    && + \pi \, i \, \mbox{Res}[f^+_n, 0] - \pi \, i \, \mbox{Res}[f^-_n,0].
\eea

Calculating the residues, we find the values of each of the $J_n$:
\bea
J_0 &=& \frac{\pi \, e^{i \, r \, (\omega + i \, \epsilon)}}{r \, (\omega + i \, \epsilon)^2} - \frac{\pi}{r \, (\omega + i \, \epsilon)^2}, \label{eqn:J0_res}\\
J_2 &=& \frac{\pi \, e^{i \, r \, (\omega + i \, \epsilon)}}{r \, (\omega + i \, \epsilon)^2} + \frac{\pi \, e^{ i \, r (\omega + i \, \epsilon)} \, [ - 2  + 2 \, i \, r \, (\omega + i \, \epsilon)]}{r^3 \, (\omega + i \, \epsilon)^4} \nonumber \\
 && + \frac{2 \, \pi}{r^3 \, (\omega + i \, \epsilon)^4}, \label{eqn:J2_res}\\
J_4 &=&\frac{\pi \, e^{i \, r \, (\omega + i \, \epsilon)}}{r \, (\omega + i \, \epsilon)^2} + \frac{4 \pi \, e^{ i \, r (\omega + i \, \epsilon)}}{r^5 \, (\omega + i \, \epsilon)^6} \, \left[ 6  - 6 \, i \, r \, (\omega + i \, \epsilon) \right. \nonumber \\
&& \left. - 3 \, r^2 \, (\omega + i \, \epsilon)^2 + i \, r^3 \, (\omega + i \, \epsilon)^3 \right] \nonumber \\
&& - \frac{24 \, \pi}{r^5 \, (\omega + i \, \epsilon)^6}. \label{eqn:J4_res}
\eea

\subsection{Frequency integration}

Now we perform the $\omega$ integration. Inserting the results
(\ref{eqn:J0_res}-\ref{eqn:J4_res}) into
(\ref{eqn:I_def}-\ref{eqn:L_def}) respectively, we see that each of
$I$, $K$ and $L$ contains a delta function, which we can extract:
\bean
I &=& \frac{1}{4 \, \pi \, r} [ \delta(T - r) - \delta(T) ],\\
K &=& \frac{1}{4 \, \pi \, r} \delta(T - r) + e^{ - r \, \epsilon } \, \int \frac{d\omega}{(2 \, \pi)^3} \, e^{ - i \omega \, (T - r) } \, F_{2a}(\omega) \\
&& + \int \frac{d\omega}{(2 \, \pi)^3}  \, e^{- i \, \omega \, T} \, F_{2b}(\omega),\\
L &=& \frac{1}{4 \, \pi \, r} \delta(T - r) + e^{ - r \, \epsilon } \, \int \frac{d\omega}{(2 \, \pi)^3} \, e^{ - i \omega \, (T - r) } \, F_{4a}(\omega) \\
&& + \int \frac{d\omega}{(2 \, \pi)^3}  \, e^{- i \, \omega \, T} \, F_{4b}(\omega),
\eean
where the new terms on the right-hand side come from the $J_n$ above, grouped by
exponential, as that is what determines the contours chosen during
integration (see Fig. \ref{fig:contours}):
\bean
F_{2a}(\omega) &=& \frac{\pi \, \omega^2 \, [ - 2  + 2 \, i \, r \, (\omega + i \, \epsilon)]}{r^3 \, (\omega + i \, \epsilon)^4},\\
F_{2b}(\omega) &=& \frac{2 \, \pi \, \omega^2}{r^3 \, (\omega + i \, \epsilon)^4},\\
F_{4a}(\omega) &=& \frac{\pi \, \omega^2}{r^5 \, (\omega + i \, \epsilon)^6} \, \left[ 24  - 24 \, i \, r \, (\omega + i \, \epsilon)  \right. \\
&& \left. - 12 \, r^2 \, (\omega + i \, \epsilon)^2 + 4 \, i \, r^3 \, (\omega + i \, \epsilon)^3 \right],\\
F_{4b}(\omega) &=& - \frac{24 \, \pi \, \omega^2}{r^5 \, (\omega + i \, \epsilon)^6}.
\eean
Now the residues are as follows (taking the $\epsilon \rightarrow 0$
limit):
\bean
\mbox{Res}\left[e^{ - i \omega \, (T - r) } \, F_{2a}(\omega) ,- i \, \epsilon \right] &=& \frac{2 \, \pi \, i \, T}{r^3},\\
\mbox{Res}\left[e^{ - i \omega \, T } \, F_{2b}(\omega) ,- i \, \epsilon \right] &=& - \frac{2 \, \pi \, i \, T}{r^3},\\
\mbox{Res}\left[e^{ - i \omega \, (T - r) } \, F_{4a}(\omega) ,- i \, \epsilon \right] &=& \frac{4 \, \pi \, i \, T^3}{r^5},\\
\mbox{Res}\left[e^{ - i \omega \, T } \, F_{4b}(\omega) ,- i \, \epsilon \right] &=& - \frac{4 \, \pi \, i \, T^3}{r^5}.
\eean
The only pole is at $\omega = - i \, \epsilon$, so if we can close the
contour in the upper half-plane, we will get zero.
\begin{itemize}
\item For $T < 0$, both the ``a'' and ``b'' integrals can be closed in
  $C_1$. Result: zero contribution.
\item For $0 < T < r$, the ``a'' integrals can be closed in $C_1$, but
  the ``b'' integrals must be closed in $C_2$. Result: ``b''
  contribution.
\item For $T > r$, both the ``a'' and ``b'' integrals must be closed
  in $C_2$. But then the ``a'' and ``b'' residues cancel out. Result:
  zero contribution.
\end{itemize}
Thus the only interesting contribution happens in the interval $0 < T
< r \Leftrightarrow t - r(\tau) < \tau < t$. In this case, the final
integrals yield
\bean
\int \frac{d\omega}{(2 \, \pi)^3} \, e^{ - i \omega \, (T - r) } \, F_{2b}(\omega) &=& - \frac{T}{2 \, \pi \, r^3},\\
\int \frac{d\omega}{(2 \, \pi)^3} \, e^{ - i \omega \, (T - r) } \, F_{4b}(\omega) &=& - \frac{T^3}{\pi \, r^5},
\eean
leading to the final result for $K$ and $L$:
\bean
I &=& \frac{1}{4 \, \pi \, r} \delta(T - r) - \frac{1}{4 \, \pi \, r} \delta(T),\\
K &=& \frac{1}{4 \, \pi \, r} \delta(T - r) - \Theta(T) \, \Theta(r - T) \, \frac{T}{2 \, \pi \, r^3},\\
L &=& \frac{1}{4 \, \pi \, r} \delta(T - r) - \Theta(T) \, \Theta(r - T) \, \frac{T^3}{\pi \, r^5}.
\eean

We use these to calculate the $I^{i \, j}$ and $I^{i \, j \, k \, l}$:
\begin{widetext}
\bea
I^{i \, j} &=& \left[ n^i \, n^j \, \left( \frac{1}{4 \, \pi \, r} \delta(T - r) - \Theta(T) \, \Theta(r - T) \, \frac{T}{2 \, \pi \, r^3} \right)  + \frac{1}{2} \, Q^{i \, j} \, \left( - \frac{1}{4 \, \pi \, r} \delta(T) + \Theta(T) \, \Theta(r - T) \, \frac{T}{2 \, \pi \, r^3} \right) \right]_{\tau}, \\
I^{i \, j \, k \, l} &=& \left[ n^i \, n^j \, n^k \, n^l \, \left( \frac{1}{4 \, \pi \, r} \delta(T - r) - \Theta(T) \, \Theta(r - T) \, \frac{T^3}{\pi \, r^5} \right) - 3\, Q^{(i \, j} \, n^k \, n^{l)} \, \Theta(T) \, \Theta(r - T) \, \left( \frac{T}{2 \, \pi \, r^3} - \frac{T^3}{\pi \, r^5} \right) \right. \nonumber\\
&& \left. + \frac{3}{8} \, Q^{(i \, j} \, Q^{k \, l)} \, \left( - \frac{1}{4 \, \pi \, r} \delta(T) + \Theta(T) \, \Theta(r - T) \, \left( \frac{T}{\pi \, r^3} - \frac{T^3}{\pi \, r^5} \right) \right) \right]_{\tau}. 
\eea

\subsection{Time integration}

The final integrations will be over the source time $\tau$. The
``crossing times'' for the two $\Theta$ functions are $\tau = t$ and
$\tau = t^r$, where $t$ is the present field time, and $t^r$ the
corresponding retarded time defined by (\ref{eqn:tret_def}). Now
taking a general function $y(\tau)$, we find that
\bean
\int_{-\infty}^{\infty} d\tau \, I_A \, y(\tau) &=& \frac{y(t_A^r)}{4 \, \pi \, r_A(t_A^r)} - \frac{y(t)}{4 \, \pi \, r_A(t)},\\
\int_{-\infty}^{\infty} d\tau \, I^{i \, j}_A \, y(\tau) &=& \left[ n_A^i \, n_A^j \, \frac{y(\tau)}{4 \, \pi \, r_A} \right]_{\tau = t_A^r} - \left[ \frac{1}{2} \, Q_A^{i \, j} \, \frac{y(\tau)}{4 \, \pi \, r_A} \right]_{\tau = t} - \int_{t_A^r}^{t} d\tau \, \left( 3 \, n_A^i \, n_A^j  -  \delta^{ij} \right) \, \frac{(t - \tau) \, y(\tau)}{4 \, \pi \, r_A(\tau)^3},\\
\int_{-\infty}^{\infty} d\tau \, I^{i \, j \, k \, l}_A \, y(\tau) &=& \left[ n_A^i \, n_A^j \, n_A^k \, n_A^l \, \frac{y(\tau)}{4 \, \pi \, r_A} \right]_{\tau = t_A^r} - \left[ \frac{3}{8} \, Q_A^{(i \, j} \, Q_A^{k \, l)} \, \frac{y(\tau)}{4 \, \pi \, r_A} \right]_{\tau = t} \\
&& + \int_{t_A^r}^{t} d\tau \, \left( - 3\, Q_A^{(i \, j} \, n_A^k \, n_A^{l)} + \frac{3}{4} \, Q_A^{(i \, j} \, Q_A^{k \, l)} \right) \, \frac{(t - \tau)}{2 \, \pi \, r_A(\tau)^3} \, y(\tau)\\
&& + \int_{t_A^r}^{t} d\tau \, \left( - n_A^i \, n_A^j \, n_A^k \, n_A^l + 3\, Q_A^{(i \, j} \, n_A^k \, n_A^{l)} - \frac{3}{8} \, Q_A^{(i \, j} \, Q_A^{k \, l)} \right) \, \frac{(t - \tau)^3}{\pi \, r_A(\tau)^5} \, y(\tau).
\eean

These can now be substituted into the general integral
(\ref{eqn:HTT_Idecomp}). We write the result as a sum of terms at the
\emph{present} field-point time $t$, the \emph{retarded} time $t_A^r$,
and \emph{interval} terms between them,
\[
H^{i \, j}_{{\rm TT}\,A} [\vec{u}] = H^{i \, j}_{{\rm TT}\,A} [\vec{u};t] + H^{i \, j}_{{\rm TT}\,A} [\vec{u};t_A^r] + H^{i \, j}_{{\rm TT}\,A} [\vec{u};t_A^r \rightarrow t],
\]

\bean
H^{i \, j}_{{\rm TT}\,A} [\vec{u};t] &=& - \frac{4\,G}{r_A(t)} \left\{ \left[ u^i \, u^j - \frac{u^2}{2} \, \delta^{ij} \right]_{t} + \left[ \frac{u^2}{2} \right]_{t} \, \frac{1}{2} \, Q_A^{i \, j} + \left[ \frac{u_k \, u_l}{2} \right]_{t} \, \frac{1}{2} \, Q_A^{k \, l} \, \delta^{ij}  \right.\\ 
& & \left. - \left[ 2 \, u_k \, u^{(i} \right]_{t} \, \frac{1}{2} \, Q_A^{j) k} + \left[ \frac{u_k \, u_l}{2} \right]_{t} \, \frac{3}{8} \, Q_A^{(i \, j} \, Q_A^{k \, l)} \right\},
\eean
\bean
H^{i \, j}_{{\rm TT}\,A} [\vec{u};t_A^r] &=& \frac{4\,G}{r_A(t_A^r)} \left\{ \left[ u^i \, u^j - \frac{u^2}{2} \,
\delta^{ij} \right]_{t_A^r} + \left[ \frac{u^2}{2} \right]_{t_A^r} \, n_A^i \, n_A^j  + \left[ \frac{u_k \,
u_l}{2} \right]_{t_A^r} \, n_A^k \, n_A^l \, \delta^{ij} \right.\\
& & \left. - \left[ 2 \, u_k \, u^{(i} \right]_{t_A^r} \, \, n_A^{j)} \, n_A^k + \left[ \frac{u_k \, u_l}{2} \right]_{t_A^r} \, n_A^i \, n_A^j \, n_A^k \, n_A^l \right\},
\eean
\bean
H^{i \, j}_{{\rm TT}\,A} [\vec{u};t_A^r \rightarrow t] &=& - 4\,G \int_{t_A^r}^{t} d\tau \, \frac{(t - \tau)}{r_A(\tau)^3} \, \left\{ \left[ \frac{u^2}{2} \right] \, \left( 3 \, n_A^i \, n_A^j  -  \delta^{ij} \right) + \left[ \frac{u_k \, u_l}{2} \right] \, \left( 3 \, n_A^k \, n_A^l  -  \delta^{k \, l} \right) \, \delta^{ij} \right. \\
&& \left. - \left[ 2 \, u_k \, u^{(i} \right] \, \left( 3 \, n_A^{j)} \, n_A^k  -  \delta^{j) \, k} \right) + \left[ \frac{u_k \, u_l}{2} \right] \, \left( 6\, Q_A^{(i \, j} \, n_A^k \, n_A^{l)} - \frac{3}{2} \, Q_A^{(i \, j} \, Q_A^{k \, l)} \right) \right\} \\
&& -  16\,G \int_{t_A^r}^{t} d\tau \, \frac{(t - \tau)^3}{r_A(\tau)^5} \, \left\{ \left[ \frac{u_k \, u_l}{2} \right] \, \left( n_A^i \, n_A^j \, n_A^k \, n_A^l - 3\, Q_A^{(i \, j} \, n_A^k \, n_A^{l)} + \frac{3}{8} \, Q_A^{(i \, j} \, Q_A^{k \, l)} \right)  \right\}.
\eean
\end{widetext}


\begin{thebibliography}{60}
\expandafter\ifx\csname natexlab\endcsname\relax\def\natexlab#1{#1}\fi
\expandafter\ifx\csname bibnamefont\endcsname\relax
  \def\bibnamefont#1{#1}\fi
\expandafter\ifx\csname bibfnamefont\endcsname\relax
  \def\bibfnamefont#1{#1}\fi
\expandafter\ifx\csname citenamefont\endcsname\relax
  \def\citenamefont#1{#1}\fi
\expandafter\ifx\csname url\endcsname\relax
  \def\url#1{\texttt{#1}}\fi
\expandafter\ifx\csname urlprefix\endcsname\relax\def\urlprefix{URL }\fi
\providecommand{\bibinfo}[2]{#2}
\providecommand{\eprint}[2][]{\url{#2}}

\bibitem[{\citenamefont{Vogt}(1992)}]{LIGO}
\bibinfo{author}{\bibfnamefont{R.}~\bibnamefont{Vogt}}, in
  \emph{\bibinfo{booktitle}{Sixth {M}arcel {G}rossman Meeting on General
  Relativity (Proceedings, Kyoto, Japan, 1991)}}, edited by
  \bibinfo{editor}{\bibfnamefont{H.}~\bibnamefont{Sato}} \bibnamefont{and}
  \bibinfo{editor}{\bibfnamefont{T.}~\bibnamefont{Nakamura}}
  (\bibinfo{publisher}{World {S}cientific}, \bibinfo{address}{Singapore},
  \bibinfo{year}{1992}), pp. \bibinfo{pages}{244--266}.

\bibitem[{\citenamefont{Abbott et~al.}(2004)}]{Abbott:2003vs}
\bibinfo{author}{\bibfnamefont{B.}~\bibnamefont{Abbott}} \bibnamefont{et~al.}
  (\bibinfo{collaboration}{LIGO Scientific}), \bibinfo{journal}{Nucl. Instrum.
  Meth.} \textbf{\bibinfo{volume}{A517}}, \bibinfo{pages}{154}
  (\bibinfo{year}{2004}), \eprint{gr-qc/0308043}.

\bibitem[{\citenamefont{Bender et~al.}(1998)}]{LISA:1998pa}
\bibinfo{author}{\bibfnamefont{P.}~\bibnamefont{Bender}} \bibnamefont{et~al.},
  \bibinfo{type}{Tech. Rep.} \bibinfo{number}{{MPQ 233}},
  \bibinfo{institution}{{Max-Planck-Institut f\"ur Quantenoptik}}
  (\bibinfo{year}{1998}), \bibinfo{note}{{URL}: {\tt
  http://www.lisa-science.org/resources/talks\\-articles/mission/prephasea.pdf%
}}.

\bibitem[{\citenamefont{Danzmann and Rudiger}(2003)}]{Danzmann:2003tv}
\bibinfo{author}{\bibfnamefont{K.}~\bibnamefont{Danzmann}} \bibnamefont{and}
  \bibinfo{author}{\bibfnamefont{A.}~\bibnamefont{Rudiger}},
  \bibinfo{journal}{Class. Quantum Grav.} \textbf{\bibinfo{volume}{20}},
  \bibinfo{pages}{S1} (\bibinfo{year}{2003}).

\bibitem[{\citenamefont{Buonanno et~al.}(2007)\citenamefont{Buonanno, Cook, and
  Pretorius}}]{Buonanno:2006ui}
\bibinfo{author}{\bibfnamefont{A.}~\bibnamefont{Buonanno}},
  \bibinfo{author}{\bibfnamefont{G.~B.} \bibnamefont{Cook}}, \bibnamefont{and}
  \bibinfo{author}{\bibfnamefont{F.}~\bibnamefont{Pretorius}},
  \bibinfo{journal}{Phys. Rev. D} \textbf{\bibinfo{volume}{75}},
  \bibinfo{pages}{124018} (\bibinfo{year}{2007}), \eprint{gr-qc/0610122}.

\bibitem[{\citenamefont{Berti et~al.}(2007)\citenamefont{Berti, Cardoso,
  Gonzalez, Sperhake, Hannam, Husa, and Br\"ugmann}}]{Berti:2007fi}
\bibinfo{author}{\bibfnamefont{E.}~\bibnamefont{Berti}},
  \bibinfo{author}{\bibfnamefont{V.}~\bibnamefont{Cardoso}},
  \bibinfo{author}{\bibfnamefont{J.~A.} \bibnamefont{Gonzalez}},
  \bibinfo{author}{\bibfnamefont{U.}~\bibnamefont{Sperhake}},
  \bibinfo{author}{\bibfnamefont{M.}~\bibnamefont{Hannam}},
  \bibinfo{author}{\bibfnamefont{S.}~\bibnamefont{Husa}}, \bibnamefont{and}
  \bibinfo{author}{\bibfnamefont{B.}~\bibnamefont{Br\"ugmann}}
  (\bibinfo{year}{2007}), \eprint{arXiv:gr-qc/0703053}.

\bibitem[{\citenamefont{Pretorius}(2005)}]{Pretorius:2005gq}
\bibinfo{author}{\bibfnamefont{F.}~\bibnamefont{Pretorius}},
  \bibinfo{journal}{Phys. Rev. Lett.} \textbf{\bibinfo{volume}{95}},
  \bibinfo{pages}{121101} (\bibinfo{year}{2005}), \eprint{gr-qc/0507014}.

\bibitem[{\citenamefont{Campanelli
  et~al.}(2006{\natexlab{a}})\citenamefont{Campanelli, Lousto, Marronetti, and
  Zlochower}}]{Campanelli:2005dd}
\bibinfo{author}{\bibfnamefont{M.}~\bibnamefont{Campanelli}},
  \bibinfo{author}{\bibfnamefont{C.~O.} \bibnamefont{Lousto}},
  \bibinfo{author}{\bibfnamefont{P.}~\bibnamefont{Marronetti}},
  \bibnamefont{and}
  \bibinfo{author}{\bibfnamefont{Y.}~\bibnamefont{Zlochower}},
  \bibinfo{journal}{Phys. Rev. Lett.} \textbf{\bibinfo{volume}{96}},
  \bibinfo{pages}{111101} (\bibinfo{year}{2006}{\natexlab{a}}),
  \eprint{gr-qc/0511048}.

\bibitem[{\citenamefont{Baker et~al.}(2006{\natexlab{a}})\citenamefont{Baker,
  Centrella, Choi, Koppitz, and van Meter}}]{Baker:2005vv}
\bibinfo{author}{\bibfnamefont{J.~G.} \bibnamefont{Baker}},
  \bibinfo{author}{\bibfnamefont{J.}~\bibnamefont{Centrella}},
  \bibinfo{author}{\bibfnamefont{D.-I.} \bibnamefont{Choi}},
  \bibinfo{author}{\bibfnamefont{M.}~\bibnamefont{Koppitz}}, \bibnamefont{and}
  \bibinfo{author}{\bibfnamefont{J.}~\bibnamefont{van Meter}},
  \bibinfo{journal}{Phys. Rev. Lett.} \textbf{\bibinfo{volume}{96}},
  \bibinfo{pages}{111102} (\bibinfo{year}{2006}{\natexlab{a}}),
  \eprint{gr-qc/0511103}.

\bibitem[{\citenamefont{Brandt and Br\"{u}gmann}(1997)}]{Brandt:1997tf}
\bibinfo{author}{\bibfnamefont{S.}~\bibnamefont{Brandt}} \bibnamefont{and}
  \bibinfo{author}{\bibfnamefont{B.}~\bibnamefont{Br\"{u}gmann}},
  \bibinfo{journal}{Phys. Rev. Lett.} \textbf{\bibinfo{volume}{78}},
  \bibinfo{pages}{3606} (\bibinfo{year}{1997}), \eprint{gr-qc/9703066}.

\bibitem[{\citenamefont{Shibata and Nakamura}(1995)}]{Shibata:1995we}
\bibinfo{author}{\bibfnamefont{M.}~\bibnamefont{Shibata}} \bibnamefont{and}
  \bibinfo{author}{\bibfnamefont{T.}~\bibnamefont{Nakamura}},
  \bibinfo{journal}{Phys. Rev. D} \textbf{\bibinfo{volume}{52}},
  \bibinfo{pages}{5428} (\bibinfo{year}{1995}).

\bibitem[{\citenamefont{Baumgarte and Shapiro}(1999)}]{Baumgarte:1998te}
\bibinfo{author}{\bibfnamefont{T.}~\bibnamefont{Baumgarte}} \bibnamefont{and}
  \bibinfo{author}{\bibfnamefont{S.}~\bibnamefont{Shapiro}},
  \bibinfo{journal}{Phys. Rev. D} \textbf{\bibinfo{volume}{59}},
  \bibinfo{pages}{024007} (\bibinfo{year}{1999}), \eprint{gr-qc/9810065}.

\bibitem[{\citenamefont{Br\"{u}gmann et~al.}(2004)\citenamefont{Br\"{u}gmann,
  Tichy, and Jansen}}]{Bruegmann:2003aw}
\bibinfo{author}{\bibfnamefont{B.}~\bibnamefont{Br\"{u}gmann}},
  \bibinfo{author}{\bibfnamefont{W.}~\bibnamefont{Tichy}}, \bibnamefont{and}
  \bibinfo{author}{\bibfnamefont{N.}~\bibnamefont{Jansen}},
  \bibinfo{journal}{Phys. Rev. Lett.} \textbf{\bibinfo{volume}{92}},
  \bibinfo{pages}{211101} (\bibinfo{year}{2004}), \eprint{gr-qc/0312112}.

\bibitem[{\citenamefont{Campanelli
  et~al.}(2006{\natexlab{b}})\citenamefont{Campanelli, Lousto, and
  Zlochower}}]{Campanelli:2006gf}
\bibinfo{author}{\bibfnamefont{M.}~\bibnamefont{Campanelli}},
  \bibinfo{author}{\bibfnamefont{C.~O.} \bibnamefont{Lousto}},
  \bibnamefont{and}
  \bibinfo{author}{\bibfnamefont{Y.}~\bibnamefont{Zlochower}},
  \bibinfo{journal}{Phys. Rev. D} \textbf{\bibinfo{volume}{73}},
  \bibinfo{pages}{061501(R)} (\bibinfo{year}{2006}{\natexlab{b}}),
  \eprint{gr-qc/0601091}.

\bibitem[{\citenamefont{Pretorius}(2006)}]{Pretorius:2006tp}
\bibinfo{author}{\bibfnamefont{F.}~\bibnamefont{Pretorius}},
  \bibinfo{journal}{Class. Quantum Grav.} \textbf{\bibinfo{volume}{23}},
  \bibinfo{pages}{S529} (\bibinfo{year}{2006}), \eprint{gr-qc/0602115}.

\bibitem[{\citenamefont{Baker et~al.}(2006{\natexlab{b}})\citenamefont{Baker,
  Centrella, Choi, Koppitz, and van Meter}}]{Baker:2006yw}
\bibinfo{author}{\bibfnamefont{J.~G.} \bibnamefont{Baker}},
  \bibinfo{author}{\bibfnamefont{J.}~\bibnamefont{Centrella}},
  \bibinfo{author}{\bibfnamefont{D.-I.} \bibnamefont{Choi}},
  \bibinfo{author}{\bibfnamefont{M.}~\bibnamefont{Koppitz}}, \bibnamefont{and}
  \bibinfo{author}{\bibfnamefont{J.}~\bibnamefont{van Meter}},
  \bibinfo{journal}{Phys. Rev. D} \textbf{\bibinfo{volume}{73}},
  \bibinfo{pages}{104002} (\bibinfo{year}{2006}{\natexlab{b}}),
  \eprint{gr-qc/0602026}.

\bibitem[{\citenamefont{Br\"{u}gmann et~al.}(2006)}]{Bruegmann:2006at}
\bibinfo{author}{\bibfnamefont{B.}~\bibnamefont{Br\"{u}gmann}}
  \bibnamefont{et~al.} (\bibinfo{year}{2006}), \eprint{gr-qc/0610128}.

\bibitem[{\citenamefont{Scheel et~al.}(2006)}]{Scheel:2006gg}
\bibinfo{author}{\bibfnamefont{M.~A.} \bibnamefont{Scheel}}
  \bibnamefont{et~al.}, \bibinfo{journal}{Phys. Rev. D}
  \textbf{\bibinfo{volume}{74}}, \bibinfo{pages}{104006}
  (\bibinfo{year}{2006}), \eprint{gr-qc/0607056}.

\bibitem[{\citenamefont{Marronetti et~al.}(2007)\citenamefont{Marronetti,
  Tichy, Br\"{u}gmann, Gonzalez, Hannam, Husa, and
  Sperhake}}]{Marronetti:2007ya}
\bibinfo{author}{\bibfnamefont{P.}~\bibnamefont{Marronetti}},
  \bibinfo{author}{\bibfnamefont{W.}~\bibnamefont{Tichy}},
  \bibinfo{author}{\bibfnamefont{B.}~\bibnamefont{Br\"{u}gmann}},
  \bibinfo{author}{\bibfnamefont{J.}~\bibnamefont{Gonzalez}},
  \bibinfo{author}{\bibfnamefont{M.}~\bibnamefont{Hannam}},
  \bibinfo{author}{\bibfnamefont{S.}~\bibnamefont{Husa}}, \bibnamefont{and}
  \bibinfo{author}{\bibfnamefont{U.}~\bibnamefont{Sperhake}},
  \bibinfo{journal}{Class. Quant. Grav.} \textbf{\bibinfo{volume}{24}},
  \bibinfo{pages}{S43} (\bibinfo{year}{2007}), \eprint{gr-qc/0701123}.

\bibitem[{\citenamefont{Tichy}(2006)}]{Tichy:2006qn}
\bibinfo{author}{\bibfnamefont{W.}~\bibnamefont{Tichy}},
  \bibinfo{journal}{Phys. Rev. D} \textbf{\bibinfo{volume}{74}},
  \bibinfo{pages}{084005} (\bibinfo{year}{2006}), \eprint{gr-qc/0609087}.

\bibitem[{\citenamefont{Pfeiffer et~al.}(2007)\citenamefont{Pfeiffer, Brown,
  Kidder, Lindblom, Lovelace, and Scheel}}]{Pfeiffer:2007yz}
\bibinfo{author}{\bibfnamefont{H.~P.} \bibnamefont{Pfeiffer}},
  \bibinfo{author}{\bibfnamefont{D.~A.} \bibnamefont{Brown}},
  \bibinfo{author}{\bibfnamefont{L.~E.} \bibnamefont{Kidder}},
  \bibinfo{author}{\bibfnamefont{L.}~\bibnamefont{Lindblom}},
  \bibinfo{author}{\bibfnamefont{G.}~\bibnamefont{Lovelace}}, \bibnamefont{and}
  \bibinfo{author}{\bibfnamefont{M.~A.} \bibnamefont{Scheel}},
  \bibinfo{journal}{Class. Quant. Grav.} pp. \bibinfo{pages}{S59--S81}
  (\bibinfo{year}{2007}), \eprint{gr-qc/0702106}.

\bibitem[{\citenamefont{Baker et~al.}(2007{\natexlab{a}})\citenamefont{Baker,
  Campanelli, Pretorius, and Zlochower}}]{Baker:2007fb}
\bibinfo{author}{\bibfnamefont{J.~G.} \bibnamefont{Baker}},
  \bibinfo{author}{\bibfnamefont{M.}~\bibnamefont{Campanelli}},
  \bibinfo{author}{\bibfnamefont{F.}~\bibnamefont{Pretorius}},
  \bibnamefont{and}
  \bibinfo{author}{\bibfnamefont{Y.}~\bibnamefont{Zlochower}},
  \bibinfo{journal}{Class. Quant. Grav.} \textbf{\bibinfo{volume}{24}},
  \bibinfo{pages}{S25} (\bibinfo{year}{2007}{\natexlab{a}}),
  \eprint{gr-qc/0701016}.

\bibitem[{\citenamefont{Thornburg et~al.}(2007)\citenamefont{Thornburg, Diener,
  Pollney, Rezzolla, Schnetter, Seidel, and Takahashi}}]{Thornburg:2007hu}
\bibinfo{author}{\bibfnamefont{J.}~\bibnamefont{Thornburg}},
  \bibinfo{author}{\bibfnamefont{P.}~\bibnamefont{Diener}},
  \bibinfo{author}{\bibfnamefont{D.}~\bibnamefont{Pollney}},
  \bibinfo{author}{\bibfnamefont{L.}~\bibnamefont{Rezzolla}},
  \bibinfo{author}{\bibfnamefont{E.}~\bibnamefont{Schnetter}},
  \bibinfo{author}{\bibfnamefont{E.}~\bibnamefont{Seidel}}, \bibnamefont{and}
  \bibinfo{author}{\bibfnamefont{R.}~\bibnamefont{Takahashi}},
  \bibinfo{journal}{Class. Quant. Grav.} \textbf{\bibinfo{volume}{24}},
  \bibinfo{pages}{3911} (\bibinfo{year}{2007}), \eprint{gr-qc/0701038}.

\bibitem[{NRw()}]{NRwaves}
\bibinfo{note}{NRwaves home page:\\{\tt
  https://gravity.psu.edu/wiki\_NRwaves}}.

\bibitem[{\citenamefont{Baker et~al.}(2007{\natexlab{b}})\citenamefont{Baker,
  McWilliams, van Meter, Centrella, Choi, Kelly, and Koppitz}}]{Baker:2006kr}
\bibinfo{author}{\bibfnamefont{J.~G.} \bibnamefont{Baker}},
  \bibinfo{author}{\bibfnamefont{S.~T.} \bibnamefont{McWilliams}},
  \bibinfo{author}{\bibfnamefont{J.~R.} \bibnamefont{van Meter}},
  \bibinfo{author}{\bibfnamefont{J.}~\bibnamefont{Centrella}},
  \bibinfo{author}{\bibfnamefont{D.-I.} \bibnamefont{Choi}},
  \bibinfo{author}{\bibfnamefont{B.~J.} \bibnamefont{Kelly}}, \bibnamefont{and}
  \bibinfo{author}{\bibfnamefont{M.}~\bibnamefont{Koppitz}},
  \bibinfo{journal}{Phys. Rev. D} \textbf{\bibinfo{volume}{75}},
  \bibinfo{pages}{124024} (\bibinfo{year}{2007}{\natexlab{b}}),
  \eprint{gr-qc/0612117}.

\bibitem[{\citenamefont{Baker et~al.}(2006{\natexlab{c}})\citenamefont{Baker,
  van Meter, McWilliams, Centrella, and Kelly}}]{Baker:2006ha}
\bibinfo{author}{\bibfnamefont{J.~G.} \bibnamefont{Baker}},
  \bibinfo{author}{\bibfnamefont{J.~R.} \bibnamefont{van Meter}},
  \bibinfo{author}{\bibfnamefont{S.~T.} \bibnamefont{McWilliams}},
  \bibinfo{author}{\bibfnamefont{J.}~\bibnamefont{Centrella}},
  \bibnamefont{and} \bibinfo{author}{\bibfnamefont{B.~J.} \bibnamefont{Kelly}}
  (\bibinfo{year}{2006}{\natexlab{c}}), \eprint{gr-qc/0612024}.

\bibitem[{\citenamefont{Campanelli}(2005)}]{Campanelli:2004zw}
\bibinfo{author}{\bibfnamefont{M.}~\bibnamefont{Campanelli}},
  \bibinfo{journal}{Class. Quant. Grav.} \textbf{\bibinfo{volume}{22}},
  \bibinfo{pages}{S387} (\bibinfo{year}{2005}), \eprint{astro-ph/0411744}.

\bibitem[{\citenamefont{Herrmann et~al.}(2006)\citenamefont{Herrmann,
  Shoemaker, and Laguna}}]{Herrmann:2006ks}
\bibinfo{author}{\bibfnamefont{F.}~\bibnamefont{Herrmann}},
  \bibinfo{author}{\bibfnamefont{D.}~\bibnamefont{Shoemaker}},
  \bibnamefont{and} \bibinfo{author}{\bibfnamefont{P.}~\bibnamefont{Laguna}}
  (\bibinfo{year}{2006}), \eprint{gr-qc/0601026}.

\bibitem[{\citenamefont{Baker et~al.}(2006{\natexlab{d}})}]{Baker:2006vn}
\bibinfo{author}{\bibfnamefont{J.~G.} \bibnamefont{Baker}}
  \bibnamefont{et~al.}, \bibinfo{journal}{Astrophys. J.}
  \textbf{\bibinfo{volume}{653}}, \bibinfo{pages}{L93}
  (\bibinfo{year}{2006}{\natexlab{d}}), \eprint{astro-ph/0603204}.

\bibitem[{\citenamefont{Campanelli
  et~al.}(2006{\natexlab{c}})\citenamefont{Campanelli, Lousto, and
  Zlochower}}]{Campanelli:2006uy}
\bibinfo{author}{\bibfnamefont{M.}~\bibnamefont{Campanelli}},
  \bibinfo{author}{\bibfnamefont{C.~O.} \bibnamefont{Lousto}},
  \bibnamefont{and}
  \bibinfo{author}{\bibfnamefont{Y.}~\bibnamefont{Zlochower}},
  \bibinfo{journal}{Phys. Rev. D} \textbf{\bibinfo{volume}{74}},
  \bibinfo{pages}{041501(R)} (\bibinfo{year}{2006}{\natexlab{c}}),
  \eprint{gr-qc/0604012}.

\bibitem[{\citenamefont{Campanelli
  et~al.}(2006{\natexlab{d}})\citenamefont{Campanelli, Lousto, and
  Zlochower}}]{Campanelli:2006fg}
\bibinfo{author}{\bibfnamefont{M.}~\bibnamefont{Campanelli}},
  \bibinfo{author}{\bibfnamefont{C.~O.} \bibnamefont{Lousto}},
  \bibnamefont{and}
  \bibinfo{author}{\bibfnamefont{Y.}~\bibnamefont{Zlochower}},
  \bibinfo{journal}{Phys. Rev. D} \textbf{\bibinfo{volume}{74}},
  \bibinfo{pages}{084023} (\bibinfo{year}{2006}{\natexlab{d}}),
  \eprint{astro-ph/0608275}.

\bibitem[{\citenamefont{Campanelli
  et~al.}(2007{\natexlab{a}})\citenamefont{Campanelli, Lousto, Zlochower,
  Krishnan, and Merritt}}]{Campanelli:2006fy}
\bibinfo{author}{\bibfnamefont{M.}~\bibnamefont{Campanelli}},
  \bibinfo{author}{\bibfnamefont{C.~O.} \bibnamefont{Lousto}},
  \bibinfo{author}{\bibfnamefont{Y.}~\bibnamefont{Zlochower}},
  \bibinfo{author}{\bibfnamefont{B.}~\bibnamefont{Krishnan}}, \bibnamefont{and}
  \bibinfo{author}{\bibfnamefont{D.}~\bibnamefont{Merritt}},
  \bibinfo{journal}{Phys. Rev. D} \textbf{\bibinfo{volume}{75}},
  \bibinfo{pages}{064030} (\bibinfo{year}{2007}{\natexlab{a}}),
  \eprint{gr-qc/0612076}.

\bibitem[{\citenamefont{Gonzalez
  et~al.}(2007{\natexlab{a}})\citenamefont{Gonzalez, Sperhake, Bruegmann,
  Hannam, and Husa}}]{Gonzalez:2006md}
\bibinfo{author}{\bibfnamefont{J.~A.} \bibnamefont{Gonzalez}},
  \bibinfo{author}{\bibfnamefont{U.}~\bibnamefont{Sperhake}},
  \bibinfo{author}{\bibfnamefont{B.}~\bibnamefont{Bruegmann}},
  \bibinfo{author}{\bibfnamefont{M.}~\bibnamefont{Hannam}}, \bibnamefont{and}
  \bibinfo{author}{\bibfnamefont{S.}~\bibnamefont{Husa}},
  \bibinfo{journal}{Phys. Rev. Lett.} \textbf{\bibinfo{volume}{98}},
  \bibinfo{pages}{091101} (\bibinfo{year}{2007}{\natexlab{a}}),
  \eprint{gr-qc/0610154}.

\bibitem[{\citenamefont{Herrmann et~al.}(2007)\citenamefont{Herrmann, Hinder,
  Shoemaker, Laguna, and Matzner}}]{Herrmann:2007ac}
\bibinfo{author}{\bibfnamefont{F.}~\bibnamefont{Herrmann}},
  \bibinfo{author}{\bibfnamefont{I.}~\bibnamefont{Hinder}},
  \bibinfo{author}{\bibfnamefont{D.}~\bibnamefont{Shoemaker}},
  \bibinfo{author}{\bibfnamefont{P.}~\bibnamefont{Laguna}}, \bibnamefont{and}
  \bibinfo{author}{\bibfnamefont{R.~A.} \bibnamefont{Matzner}}
  (\bibinfo{year}{2007}), \eprint{gr-qc/0701143}.

\bibitem[{\citenamefont{Campanelli
  et~al.}(2007{\natexlab{b}})\citenamefont{Campanelli, Lousto, Zlochower, and
  Merritt}}]{Campanelli:2007ew}
\bibinfo{author}{\bibfnamefont{M.}~\bibnamefont{Campanelli}},
  \bibinfo{author}{\bibfnamefont{C.~O.} \bibnamefont{Lousto}},
  \bibinfo{author}{\bibfnamefont{Y.}~\bibnamefont{Zlochower}},
  \bibnamefont{and} \bibinfo{author}{\bibfnamefont{D.}~\bibnamefont{Merritt}},
  \textbf{\bibinfo{volume}{659}}, \bibinfo{pages}{L5}
  (\bibinfo{year}{2007}{\natexlab{b}}), \bibinfo{note}{revised version has very
  different numbers/formulae}, \eprint{gr-qc/0701164}.

\bibitem[{\citenamefont{Koppitz et~al.}(2007)\citenamefont{Koppitz, Pollney,
  Reisswig, Rezzolla, thornburg, Diener, and Schnetter}}]{Koppitz:2007ev}
\bibinfo{author}{\bibfnamefont{M.}~\bibnamefont{Koppitz}},
  \bibinfo{author}{\bibfnamefont{D.}~\bibnamefont{Pollney}},
  \bibinfo{author}{\bibfnamefont{C.}~\bibnamefont{Reisswig}},
  \bibinfo{author}{\bibfnamefont{L.}~\bibnamefont{Rezzolla}},
  \bibinfo{author}{\bibfnamefont{J.}~\bibnamefont{thornburg}},
  \bibinfo{author}{\bibfnamefont{P.}~\bibnamefont{Diener}}, \bibnamefont{and}
  \bibinfo{author}{\bibfnamefont{E.}~\bibnamefont{Schnetter}},
  \bibinfo{journal}{Phys. Rev. Lett.} \textbf{\bibinfo{volume}{99}},
  \bibinfo{pages}{041102} (\bibinfo{year}{2007}), \eprint{gr-qc/0701163}.

\bibitem[{\citenamefont{Gonzalez
  et~al.}(2007{\natexlab{b}})\citenamefont{Gonzalez, Hannam, Sperhake,
  Br\"{u}gmann, and Husa}}]{Gonzalez:2007hi}
\bibinfo{author}{\bibfnamefont{J.~A.} \bibnamefont{Gonzalez}},
  \bibinfo{author}{\bibfnamefont{M.~D.} \bibnamefont{Hannam}},
  \bibinfo{author}{\bibfnamefont{U.}~\bibnamefont{Sperhake}},
  \bibinfo{author}{\bibfnamefont{B.}~\bibnamefont{Br\"{u}gmann}},
  \bibnamefont{and} \bibinfo{author}{\bibfnamefont{S.}~\bibnamefont{Husa}},
  \bibinfo{journal}{Phys. Rev. Lett.} \textbf{\bibinfo{volume}{98}},
  \bibinfo{pages}{231101} (\bibinfo{year}{2007}{\natexlab{b}}),
  \eprint{gr-qc/0702052}.

\bibitem[{\citenamefont{Choi et~al.}(2007)}]{Choi:2007eu}
\bibinfo{author}{\bibfnamefont{D.-I.} \bibnamefont{Choi}} \bibnamefont{et~al.}
  (\bibinfo{year}{2007}), \eprint{gr-qc/0702016}.

\bibitem[{\citenamefont{Baker et~al.}(2007{\natexlab{c}})}]{Baker:2007gi}
\bibinfo{author}{\bibfnamefont{J.~G.} \bibnamefont{Baker}} \bibnamefont{et~al.}
  (\bibinfo{year}{2007}{\natexlab{c}}), \eprint{astro-ph/0702390}.

\bibitem[{\citenamefont{Pretorius and Khurana}(2007)}]{Pretorius:2007jn}
\bibinfo{author}{\bibfnamefont{F.}~\bibnamefont{Pretorius}} \bibnamefont{and}
  \bibinfo{author}{\bibfnamefont{D.}~\bibnamefont{Khurana}},
  \bibinfo{journal}{Class. Quant. Grav.} \textbf{\bibinfo{volume}{24}},
  \bibinfo{pages}{S83} (\bibinfo{year}{2007}), \eprint{gr-qc/0702084}.

\bibitem[{\citenamefont{Campanelli
  et~al.}(2007{\natexlab{c}})\citenamefont{Campanelli, Lousto, Zlochower, and
  Merritt}}]{Campanelli:2007cg}
\bibinfo{author}{\bibfnamefont{M.}~\bibnamefont{Campanelli}},
  \bibinfo{author}{\bibfnamefont{C.~O.} \bibnamefont{Lousto}},
  \bibinfo{author}{\bibfnamefont{Y.}~\bibnamefont{Zlochower}},
  \bibnamefont{and} \bibinfo{author}{\bibfnamefont{D.}~\bibnamefont{Merritt}},
  \bibinfo{journal}{Phys. Rev. Lett.} \textbf{\bibinfo{volume}{98}},
  \bibinfo{pages}{231102} (\bibinfo{year}{2007}{\natexlab{c}}),
  \eprint{arXiv:gr-qc/0702133}.

\bibitem[{\citenamefont{Tichy and Marronetti}(2007)}]{Tichy:2007hk}
\bibinfo{author}{\bibfnamefont{W.}~\bibnamefont{Tichy}} \bibnamefont{and}
  \bibinfo{author}{\bibfnamefont{P.}~\bibnamefont{Marronetti}}
  (\bibinfo{year}{2007}), \eprint{gr-qc/0703075}.

\bibitem[{\citenamefont{Sch\"{a}fer}(1985)}]{Schafer:1986rd}
\bibinfo{author}{\bibfnamefont{G.}~\bibnamefont{Sch\"{a}fer}},
  \bibinfo{journal}{Ann. Phys.} \textbf{\bibinfo{volume}{161}},
  \bibinfo{pages}{81} (\bibinfo{year}{1985}).

\bibitem[{\citenamefont{Jaranowski and Sch\"{a}fer}(1998)}]{Jaranowski:1997ky}
\bibinfo{author}{\bibfnamefont{P.}~\bibnamefont{Jaranowski}} \bibnamefont{and}
  \bibinfo{author}{\bibfnamefont{G.}~\bibnamefont{Sch\"{a}fer}},
  \bibinfo{journal}{Phys. Rev. D} \textbf{\bibinfo{volume}{57}},
  \bibinfo{pages}{7274} (\bibinfo{year}{1998}), \bibinfo{note}{errata: Phys.
  Rev. D {\bf 63}, 029902(E) (2000)}, \eprint{gr-qc/9712075}.

\bibitem[{\citenamefont{Tichy et~al.}(2003{\natexlab{a}})\citenamefont{Tichy,
  Br\"{u}gmann, Campanelli, and Diener}}]{Tichy:2002ec}
\bibinfo{author}{\bibfnamefont{W.}~\bibnamefont{Tichy}},
  \bibinfo{author}{\bibfnamefont{B.}~\bibnamefont{Br\"{u}gmann}},
  \bibinfo{author}{\bibfnamefont{M.}~\bibnamefont{Campanelli}},
  \bibnamefont{and} \bibinfo{author}{\bibfnamefont{P.}~\bibnamefont{Diener}},
  \bibinfo{journal}{Phys. Rev. D} \textbf{\bibinfo{volume}{67}},
  \bibinfo{pages}{064008} (\bibinfo{year}{2003}{\natexlab{a}}),
  \eprint{gr-qc/0207011}.

\bibitem[{Cac()}]{Cactus_webpage}
\bibinfo{note}{{\tt Cactus} Compuational Toolkit, {\tt
  http://www.cactuscode.org}}.

\bibitem[{\citenamefont{Ohta et~al.}(1974)\citenamefont{Ohta, Okamura, Kimura,
  and Hiida}}]{Ohta:1974kp}
\bibinfo{author}{\bibfnamefont{T.}~\bibnamefont{Ohta}},
  \bibinfo{author}{\bibfnamefont{H.}~\bibnamefont{Okamura}},
  \bibinfo{author}{\bibfnamefont{T.}~\bibnamefont{Kimura}}, \bibnamefont{and}
  \bibinfo{author}{\bibfnamefont{K.}~\bibnamefont{Hiida}},
  \bibinfo{journal}{Prog. Theor. Phys.} \textbf{\bibinfo{volume}{51}},
  \bibinfo{pages}{1598} (\bibinfo{year}{1974}).

\bibitem[{\citenamefont{Fock}(1964)}]{Fock}
\bibinfo{author}{\bibfnamefont{V.~A.} \bibnamefont{Fock}},
  \emph{\bibinfo{title}{The Theory of Space, Time and Gravitation, 2nd ed.}}
  (\bibinfo{publisher}{Pergamon Press}, \bibinfo{year}{1964}).

\bibitem[{\citenamefont{Tichy et~al.}(2003{\natexlab{b}})\citenamefont{Tichy,
  Br\"ugmann, and Laguna}}]{Tichy:2003zg}
\bibinfo{author}{\bibfnamefont{W.}~\bibnamefont{Tichy}},
  \bibinfo{author}{\bibfnamefont{B.}~\bibnamefont{Br\"ugmann}},
  \bibnamefont{and} \bibinfo{author}{\bibfnamefont{P.}~\bibnamefont{Laguna}},
  \bibinfo{journal}{Phys. Rev. D} \textbf{\bibinfo{volume}{68}},
  \bibinfo{pages}{064008} (\bibinfo{year}{2003}{\natexlab{b}}),
  \eprint{gr-qc/0306020}.

\bibitem[{\citenamefont{Tichy and Br{\"u}gmann}(2004)}]{Tichy:2003qi}
\bibinfo{author}{\bibfnamefont{W.}~\bibnamefont{Tichy}} \bibnamefont{and}
  \bibinfo{author}{\bibfnamefont{B.}~\bibnamefont{Br{\"u}gmann}},
  \bibinfo{journal}{Phys. Rev. D} \textbf{\bibinfo{volume}{69}},
  \bibinfo{pages}{024006} (\bibinfo{year}{2004}), \eprint{gr-qc/0307027}.

\bibitem[{\citenamefont{Ansorg et~al.}(2004)\citenamefont{Ansorg, Br\"ugmann,
  and Tichy}}]{Ansorg:2004ds}
\bibinfo{author}{\bibfnamefont{M.}~\bibnamefont{Ansorg}},
  \bibinfo{author}{\bibfnamefont{B.}~\bibnamefont{Br\"ugmann}},
  \bibnamefont{and} \bibinfo{author}{\bibfnamefont{W.}~\bibnamefont{Tichy}},
  \bibinfo{journal}{Phys. Rev. D} \textbf{\bibinfo{volume}{70}},
  \bibinfo{pages}{064011} (\bibinfo{year}{2004}), \eprint{gr-qc/0404056}.

\bibitem[{\citenamefont{Finn and Chernoff}(1993)}]{Finn:1992xs}
\bibinfo{author}{\bibfnamefont{L.~S.} \bibnamefont{Finn}} \bibnamefont{and}
  \bibinfo{author}{\bibfnamefont{D.~F.} \bibnamefont{Chernoff}},
  \bibinfo{journal}{Phys. Rev. D} \textbf{\bibinfo{volume}{47}},
  \bibinfo{pages}{2198} (\bibinfo{year}{1993}), \eprint{gr-qc/9301003}.

\bibitem[{\citenamefont{Cutler et~al.}(1993)}]{Cutler:1993tc}
\bibinfo{author}{\bibfnamefont{C.}~\bibnamefont{Cutler}} \bibnamefont{et~al.},
  \bibinfo{journal}{Phys. Rev. Lett.} \textbf{\bibinfo{volume}{70}},
  \bibinfo{pages}{2984} (\bibinfo{year}{1993}), \eprint{astro-ph/9208005}.

\bibitem[{\citenamefont{Tichy et~al.}(2000)\citenamefont{Tichy, Flanagan, and
  Poisson}}]{Tichy:1999pv}
\bibinfo{author}{\bibfnamefont{W.}~\bibnamefont{Tichy}},
  \bibinfo{author}{\bibfnamefont{E.~E.} \bibnamefont{Flanagan}},
  \bibnamefont{and} \bibinfo{author}{\bibfnamefont{E.}~\bibnamefont{Poisson}},
  \bibinfo{journal}{Phys. Rev. D} \textbf{\bibinfo{volume}{61}},
  \bibinfo{pages}{104015} (\bibinfo{year}{2000}), \eprint{gr-qc/9912075}.

\bibitem[{\citenamefont{Sch\"{a}fer and Wex}(1993)}]{SchaferWex_1993}
\bibinfo{author}{\bibfnamefont{G.}~\bibnamefont{Sch\"{a}fer}} \bibnamefont{and}
  \bibinfo{author}{\bibfnamefont{N.}~\bibnamefont{Wex}},
  \bibinfo{journal}{Phys. Lett. A} \textbf{\bibinfo{volume}{174}},
  \bibinfo{pages}{196} (\bibinfo{year}{1993}).

\bibitem[{\citenamefont{Memmesheimer et~al.}(2004)\citenamefont{Memmesheimer,
  Gopakumar, and Sch\"{a}fer}}]{Memmesheimer:2004cv}
\bibinfo{author}{\bibfnamefont{R.-M.} \bibnamefont{Memmesheimer}},
  \bibinfo{author}{\bibfnamefont{A.}~\bibnamefont{Gopakumar}},
  \bibnamefont{and}
  \bibinfo{author}{\bibfnamefont{G.}~\bibnamefont{Sch\"{a}fer}},
  \bibinfo{journal}{Phys. Rev. D} \textbf{\bibinfo{volume}{70}},
  \bibinfo{pages}{104011} (\bibinfo{year}{2004}), \eprint{gr-qc/0407049}.

\bibitem[{\citenamefont{Yunes et~al.}(2006)\citenamefont{Yunes, Tichy, Owen,
  and Br\"{u}gmann}}]{Yunes:2005nn}
\bibinfo{author}{\bibfnamefont{N.}~\bibnamefont{Yunes}},
  \bibinfo{author}{\bibfnamefont{W.}~\bibnamefont{Tichy}},
  \bibinfo{author}{\bibfnamefont{B.~J.} \bibnamefont{Owen}}, \bibnamefont{and}
  \bibinfo{author}{\bibfnamefont{B.}~\bibnamefont{Br\"{u}gmann}},
  \bibinfo{journal}{Phys. Rev. D} \textbf{\bibinfo{volume}{74}},
  \bibinfo{pages}{104011} (\bibinfo{year}{2006}), \eprint{gr-qc/0503011}.

\bibitem[{\citenamefont{Yunes and Tichy}(2006)}]{Yunes:2006iw}
\bibinfo{author}{\bibfnamefont{N.}~\bibnamefont{Yunes}} \bibnamefont{and}
  \bibinfo{author}{\bibfnamefont{W.}~\bibnamefont{Tichy}},
  \bibinfo{journal}{Phys. Rev. D} \textbf{\bibinfo{volume}{74}},
  \bibinfo{pages}{064013} (\bibinfo{year}{2006}), \eprint{gr-qc/0601046}.

\bibitem[{\citenamefont{Blanchet}(1996)}]{Blanchet:1996wx}
\bibinfo{author}{\bibfnamefont{L.}~\bibnamefont{Blanchet}},
  \bibinfo{journal}{Phys. Rev. D} \textbf{\bibinfo{volume}{54}},
  \bibinfo{pages}{1417} (\bibinfo{year}{1996}), \eprint{gr-qc/9603048}.

\bibitem[{\citenamefont{Damour et~al.}(2005)\citenamefont{Damour, Gopakumar,
  and Iyer}}]{Damour:2004bz}
\bibinfo{author}{\bibfnamefont{T.}~\bibnamefont{Damour}},
  \bibinfo{author}{\bibfnamefont{A.}~\bibnamefont{Gopakumar}},
  \bibnamefont{and} \bibinfo{author}{\bibfnamefont{B.~R.} \bibnamefont{Iyer}},
  \bibinfo{journal}{Phys. Rev. D} \textbf{\bibinfo{volume}{70}},
  \bibinfo{pages}{064028} (\bibinfo{year}{2005}), \eprint{gr-qc/0404128}.

\end{thebibliography}
\end{document}